\documentclass{vldb}
\usepackage{graphicx}
\usepackage{xspace}
\usepackage{balance}  
\usepackage{algorithm, algorithmicx, algpseudocode}
\usepackage{xcolor,colortbl}
\usepackage{subcaption}
\usepackage{url}
\usepackage{tikz,pgfplots}
\pgfplotsset{compat=1.14}
\usetikzlibrary{patterns}

\newcommand{\protocol}{TFCommit\xspace}
\newcommand{\vo}{$\mathcal{VO}$\xspace}

\newcommand{\system}{Fides\xspace}
\definecolor{LightCyan}{rgb}{0.88,1,1}
\definecolor{Gray}{gray}{0.9}
\begin{document}

% ****************** TITLE ****************************************

\title{
\system: Managing Data on Untrusted Infrastructure}

\numberofauthors{1}

\author{
\alignauthor
Sujaya Maiyya~~~~~~~~Danny Hyun Bum Cho~~~~~~~Divyakant Agrawal~~~~~~~~Amr El Abbadi\\
       \affaddr{UC Santa Barbara} 
       \email{\{sujaya\_maiyya, hyunbumcho, divyagrawal, elabbadi\}@ucsb.edu}
}

\maketitle
\abstract
    Significant amounts of data are currently
being stored and
managed on third-party servers. It is impractical
for many small scale enterprises to own their
private datacenters, hence renting third-party servers 
is a viable solution for such
businesses. But the increasing number 
of malicious attacks, both internal and external, 
as well as buggy software on third-party servers 
is causing clients to lose their trust
in these external infrastructures. While small
enterprises cannot avoid using external infrastructures, 
they need the right set of protocols to manage
their data on untrusted infrastructures.
In this paper, we propose \textbf{\textit{\protocol}},
a novel atomic commitment protocol that executes transactions
on data stored across multiple untrusted servers. To
our knowledge, \protocol is the first atomic commitment
protocol to execute transactions in an untrusted environment
without using expensive Byzantine replication.
Using \protocol, we propose
an {\em auditable} data management system, \textbf{\textit{\system}},
residing completely on untrustworthy infrastructure.
As an auditable system, \system guarantees
the detection of potentially malicious failures occurring on untrusted servers
using tamper-resistant logs with the support of cryptographic techniques.
The experimental evaluation demonstrates the scalability and
the relatively low overhead of our approach that allows
executing transactions on untrusted infrastructure.
%\end{abstract}

\section{Introduction}
\label{sec:intro}
A fundamental problem in distributed data management
is to ensure the atomic and correct execution of 
transactions. Any transaction that updates data 
stored across multiple
servers needs to be executed atomically, i.e., either
all the operations of the transaction
are executed or none of them are executed. 
This problem has been solved using commitment protocols, such as 
Two Phase Commit (2PC)~\cite{gray1978notes}. Traditionally,
the infrastructure, and hence the servers storing the
data, were considered trustworthy. A
standard assumption
was that if a server failed, it would
simply crash;
and unless a server failed, 
it executed the 
designated protocol correctly.
%Such trust assumptions may not hold true anymore.
% 2PC assumes 
% that all sites are trustworthy; a transaction is
% committed only if all servers agreed to commit,
% and aborted when at least one server either decides to
% abort or fails by crashing. Furthermore, if a server
% decides to commit, it will consistently inform the 
% same decision to all the servers.

The recent advent of cloud computing and the rise of
blockchain systems are dramatically changing the
trust assumptions about the underlying infrastructure. 
In a cloud environment, clients store their data on third-party
servers, located on one or more data centers, 
and they execute transactions on the data. 
The servers hosted in the data centers are 
vulnerable to external attacks or software bugs that can
potentially expose a client's critical data to a malign agent
(e.g., credit details exposed in Equifax data breach \cite{equifax}, breaches to Amazon S3 buckets~\cite{amazonAttack}). 
Further, a server may intentionally decide not to follow 
the protocol execution, either to improve its performance 
or for any other self-interest (e.g., the next big cyber threat is speculated to be intentional data manipulation\cite{cyberThreat}). 

The increasing popularity of blockchain is also exposing
the challenges of storing data on non-trustworthy 
infrastructures.
Applications such as
supply chain management~\cite{korpela2017digital} execute
transactions on data repositories
maintained by multiple administrative domains
that mutually distrust each other. 
Open permissionless blockchains such as Bitcoin
\cite{nakamoto2008bitcoin} use computationally
expensive mining, whereas closed
permissioned blockchains such as Hyperledger
Fabric~\cite{androulaki2018hyperledger} use
byzantine consensus protocols to
tolerate maliciously failing servers. Blockchains
resort to expensive protocols that tolerate
malicious failures because for many applications,
both the underlying infrastructure and the 
participating entities are untrusted.

The challenge of malicious untrustworthy
infrastructure has been extensively studied by
the cryptographic and security communities (e.g.,
Pinocchio~\cite{parno2013pinocchio} that verifies
outsourced computing) as well as
in the distributed systems community,
originally introduced by Lamport in the famous 
Byzantine Agreement Protocol~\cite{lamport1982byzantine}.
One main motivation for the protocol was to ensure continuous 
service availability in Replicated State Machines even 
in the presence of malicious failures.
 
In most existing databases, the prevalent approach 
to tolerate malicious
failures is by replicating either the whole database
or the transaction manager~\cite{garcia1986applications, gashi2004designing, vandiver2007tolerating, zhao2007byzantine, al2017chainspace}.
Practical Byzantine Fault Tolerance (PBFT)~\cite{castro1999practical} by Castro and Liskov
has become the predominant replication protocol
used in designing data management systems residing on
untrusted or byzantine infrastructure. 
These systems
provide fault-tolerance in that the system makes
progress in spite of byzantine failures; the replication masks 
these failures and ensures that non-faulty processes
always observe correct and reliable information. \textit{Fault
tolerance is guaranteed only if at most
one third of the replicas are faulty~\cite{bracha1985asynchronous}}. 

In a relatively open and heterogeneous environment 
knowing the number of faulty servers -- let alone placing a 
bound on them -- is unrealistic. In such settings,
an alternate approach to tolerate malicious
failures is \textit{fault-detection} which can be
achieved using
\textit{auditability}. Fault
detection imposes no bound on the number of
faulty servers -- any
server can fail maliciously but the failures are always detected 
as they are not masked from the correct servers;
detection requires only one server to be correct at any given time. To guarantee fault detection through audits, tamper-proof logs have been 
proposed and widely used in 
systems such as PeerReview~\cite{haeberlen2007peerreview} 
and CATS~\cite{yumerefendi2007strong}.

Motivated by the need to develop a fault-detection based
data management system, we make two major 
propositions in this paper.
First, we develop a data 
management system, \textit{\textbf{\system}}\footnote{\textit{\system} 
is the Roman Goddess of trust and good faith.}, consisting 
of untrusted servers that may suffer arbitrary failures
in all the layers of a typical database, i.e.,
the transaction execution layer, the distributed atomic
commitment layer, and the datastore layer. Second,
we propose a novel atomic commit protocol --\textit{\textbf{TrustFree 
Commitment}} (\protocol) -- an integral component of \system
that commits distributed 
transactions across untrusted servers while providing 
auditable guarantees. To our knowledge, \protocol
is the first to solve the distributed atomic commitment 
problem in an untrusted infrastructure without using expensive
byzantine replication protocols. Although we present \system
with \protocol as an integral component, \protocol can be
disintegrated from \system and used in any
other design of a trust-free data management.

With detection being the focus rather 
than tolerance of malicious failures, \system 
precisely identifies the point in the execution
history at which a fault occurred, as well
as the servers that acted malicious.
These guarantees provide two fold benefits: i) A 
malicious fault by a database server is eventually
detected and undeniably linked to the malicious server, and
ii) A benign server can always defend
itself against falsified accusations. 
By providing auditabiity, \system
incentivises a server not to act 
maliciously. Furthermore, by designing a
stand-alone commit protocol,
\protocol,
that leverages cryptography, we take the first step 
towards developing a full-fledged 
data management system that fully resides in untrusted
infrastructures. We believe it is critical to start with 
a strong and solid atomic commitment building block that can
be expanded to include fault tolerance and other components
of a transaction management hierarchy.

Section~\ref{sec:background} provides the necessary 
background used in developing a trust-free data management
system. Section~\ref{sec:sys_model} discusses
the architecture, system,
and failure models of \system.
Section~\ref{sec:protocol} describes the auditable
transaction model in \system and also introduces \protocol.
Section~\ref{sec:failures} provides a few
failure
examples and their detection.
Experimental evaluation of
\protocol is presented in Section~\ref{sec:eval},
followed by related work
in Section~\ref{sec:related_work}.
Section~\ref{sec:con} concludes the paper.

\section{Cryptographic Preliminaries}
\label{sec:background}

Developing a data management system built on
untrusted infrastructure
relies heavily on many cryptographic tools. In
this section, we provide the necessary cryptographic
techniques used throughout the paper.

\subsection{Digital Signatures}

A digital signature, similar to an actual signature, 
authenticates messages. A public-key
signature~\cite{rivest1978method} consists of a public
key, $p_k$, which is known to all participants,
and a secret key, $s_k$, known only to the message 
author. The author, $\mathcal{A}$, signs message 
$m$ using her secret key $s_k$. Given the message $m$
and the signature, any receiver can verify whether the author
$\mathcal{A}$ sent the message $m$ by decrypting the signature 
using $\mathcal{A}$'s public key $p_k$. 
Public-key signature schemes are used to
prevent forgery as
it is computationally infeasible for
author $\mathcal{B}$ to sign a message with author 
$\mathcal{A}$'s signature.

\subsection{Collective Signing}
\label{sub:cosi}

Multisignature (multisig) is a form of digital signature that 
allows more than one user to sign a single record. Multisigs,
such as  Schnorr
Multisignature~\cite{schnorr1991efficient},
provide additional authenticity and security compared with 
single user's signature. Collective 
Signing (CoSi)~\cite{syta2016keeping},
an optimization of Schnorr Multisigs,
allows a \textit{leader} to produce a record
which then can be publicly validated and 
signed by a group of \textit{witnesses}. 
CoSi requires two rounds of communication
to produce a \textit{collective signature} (co-sign)
with the size and verification cost of a single signature.
Figure~\ref{fig:cosi} represents the phases
of CoSi where $L$ is the leader and $1,2,..,N$
are the witnesses.
The phases of CoSi are:

\begin{figure}[t!]
\centering
\includegraphics[scale=0.32]{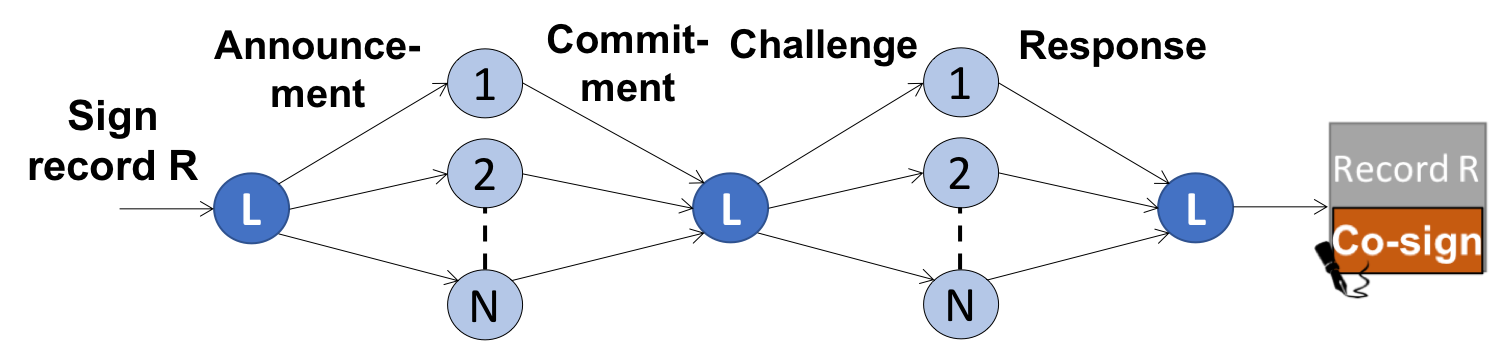}
\caption{Collective Signing.}
\vspace{-1em}
\label{fig:cosi}
\end{figure}

\textbf{Announcement}:  The leader \textit{announces}
    the beginning of a new round to
    all the witnesses and sends the record $R$ to be 
    collectively signed.
    
\textbf{Commitment}: Each witness, in response,
    picks a random secret, which is used to compute the Schnorr commitment,
    $x_{sch}$.
    The witness then sends the commitment to the leader.
    
\textbf{Challenge}: The leader aggregates all the 
    commits, $X = \sum x_{sch}$ and computes a Schnorr challenge,
    $ch = hash(X|R)$. The leader then 
    broadcasts the challenge to all the witnesses.
    
\textbf{Response}: Each witness validates the record
    before computing a Schnorr-response, $r_{sch}$, using 
    the challenge and its secret key.
    The leader collects and aggregates all the responses to 
    finally produce a Schnorr multisignature.
    
\vspace{0.5em}
The collective signature provides a \textbf{proof} that the record 
is produced by the leader and that all the witnesses signed it
only after a successful validation. Anyone with the public keys of all
the involved servers can verify the co-sign and 
the verification cost is the same as verifying 
a single signature. An invalid record will not produce
enough responses to prove the authenticity of the record.
We refer to the original work~\cite{syta2016keeping} for a detailed
discussion of the protocol.

\subsection{Merkle Hash Tree}
\label{sub:mht}
\begin{figure}[th!]
\centering
\includegraphics[scale=0.3]{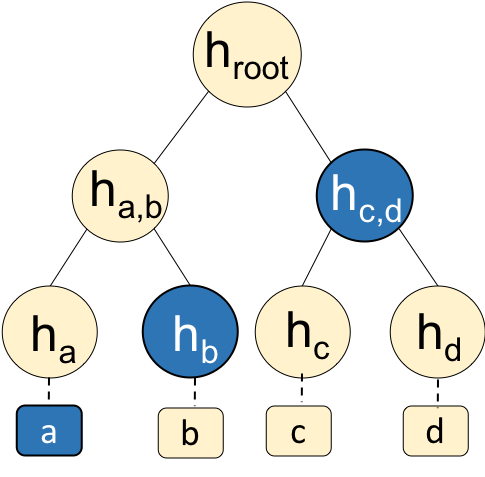}
\caption{Merkle Hash Tree example.}
\vspace{-1em}
\label{fig:mht_ex}
\end{figure}
A merkle hash tree (MHT)~\cite{merkle1989certified}
is a binary tree with each leaf node
labeled with the hash of a data item and each internal node
labeled with the hash of the concatenated labels of its
children. Figure~\ref{fig:mht_ex}
shows an example of a MHT. The hash functions, $h$, used in MHTs
are \textit{one way hash functions} i.e., for a given input $x$, 
$h(x) = y$, such that, given $y$ and $h$, it is computationally 
infeasible to obtain $x$. The hash function $h$ must also be
collision-free, i.e.,
it is highly unlikely to have two distinct inputs $x$ and $z$ that
satisfies $h(x) = h(z)$. Any such hash function can be used
to construct a MHT.

\textbf{Data Authentication Using MHTs:}
MHTs are used 
to authenticate a set of data values~\cite{merkle1989certified}
by requiring the prover, say Alice, to publicly share the 
root of the MHT, $h_{root}$, whose leave form the data set.
To authenticate a single
data value, all that a verifier, say Bob, needs from Alice is a 
\textit{Verification Object} (\vo) consisting of
all the sibling nodes along the path from the data value to 
the root. The highlighted nodes in Figure~\ref{fig:mht_ex} 
form the verification object for data
item $a$, $\mathcal{VO}(a)$, which is of size $\log_2n$. 
To authenticate data item $a$, Alice
generates the $\mathcal{VO}(a)$, and provides the value
of $a$ and $\mathcal{VO}(a)$ to Bob.
Given the
value of $a$, Bob computes $h(a)$ and uses $h_b$ from
$\mathcal{VO}(a)$ to compute $h_{a,b} = h( h(a) | h(b))$ i.e., the hash of 
 $h(a)$ concatenated with $h(b)$. Finally, using
$h_{a,b}$ and $h_{c,d}$ 
sent in the $\mathcal{VO}(a)$, Bob computes the root, $h_{a,b,c,d} = 
(h_{a,b} | h_{c,d})$.  Bob then compares the computed 
root, $h_{a,b,c,d}$, with the root publicly shared by Alice 
$h_{root}$. Assuming the use of a collision free hash function 
($h(a_1) \neq h(a_2)$ where $a_1 \neq a_2$), it would be computationally infeasible for
Alice to tamper with $a$'s value such
that the $h_{root}$ published by Alice matches the root
computed by Bob using the verification object.
\section{Fides Architecture}
\label{sec:sys_model}

\system is a data management system built on untrusted
infrastructure.
This section lays the premise for \system
by presenting the system model, the failure model,
and the audit 
mechanism of \system.

\vspace{-0.5em}
\subsection{System Model}
\system is a distributed database of
multiple servers; the data is partitioned 
into multiple shards and
distributed on these servers (perhaps 
provisioned by different providers). Shards consist
of a set of data items, each with
a unique identifier. The system 
assumes neither the servers nor the clients to be 
trustworthy and can behave arbitrarily. Servers 
and clients are uniquely identifiable
using their public keys and are aware of all the other servers 
in the system. All message exchanges (client-server or 
server-server) are digitally signed 
by the sender and verified by the receiver.

The clients interact with the data via transactions
consisting of read and write operations.
% As in the case of a traditional DBMS model, 
% transactions in isolation are assumed to be correct
% and hence, the serializable execution of multiple
% transactions is also deemed to be correct.
% 
The data can be either
single-versioned or multi-versioned with each 
committed transaction generating a new version. Every
data item has an associated read timestamp
$r_{ts}$ and a write timestamp $w_{ts}$,
indicating the timestamp of the last transaction 
that read and wrote the item, respectively. 
When a transaction commits, it updates the timestamps
of the accessed data items.

\begin{figure}[t!]
\centering
\includegraphics[scale=0.35]{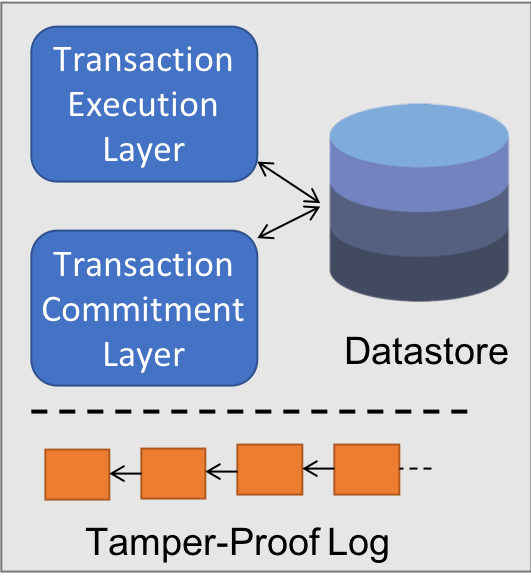}
\caption{Components of a database server.}
\label{fig:sys_model}
\vspace{-1em}
\end{figure}

We choose a simplified design for a database server to
minimize the potential for failure. As indicated in
Figure~\ref{fig:sys_model}, each database
server is composed of four components: an
\textit{execution layer} to perform transactional reads
and writes; a \textit{commitment layer} to atomically
(i.e., all servers either \textsf{commit} or 
\textsf{abort} a transaction)
terminate transactions; a \textit{datastore} where the
data shards are stored; and a \textit{tamper-proof
log}.

As individual servers are not trusted, we
replace the local transaction logs used in
traditional protocols such as 
Aries~\cite{mohan1992aries}
with a \textit{globally replicated tamper-proof
log} (this approach is inspired by blockchain).
The log -- a linked-list of transaction \textit{blocks}
linked using cryptographic hash pointers --
guarantees immutability.
Global replication of the log guarantees that even
if a subset (but not all) of the servers collude to
tamper the log, the 
transaction history is persistent.

\subsection{Failure model}
\label{sec:failure_model}

In \system, a server that fails maliciously
can behave arbitrarily i.e., send arbitrary
messages, drop messages, or corrupt the data it 
stores. \system assumes that each server and
client is computationally bounded and
is incapable of violating any cryptographic primitives
such as forging
digital signatures or breaking one-way hash functions -- the
operations that typically require brute force techniques.

Let $n$ be the total number of servers and $f$
the maximum number of faulty servers. 
\system tolerates up to $n-1$ faulty servers, i.e.,
$n>f$.
To detect failures, \system requires at least one
server to be correct and failure-free (free of malicious, 
crash, or network partition failures) at a given time. This
implies that the correct set of servers are not static and can
vary over time.
This failure model is motivated by Dolev 
and Strong's~\cite{dolev1983authenticated}
protocol where the unforgeability of digital signatures
allows tolerating up to $n$-1 failures
rather than at most $\frac{1}{3}n$ malicious failures without digital signatures.

An individual server, comprising of four components
as shown in Figure~\ref{fig:sys_model},
can fail at one or more of the
components. A fault in the \textit{execution layer} can 
return incorrect 
values; in the \textit{commit layer} can violate 
transaction
atomicity; in the \textit{datastore} can corrupt the 
stored data
values; and in the \textit{log} can omit
or reorder the transaction history. We discuss
these faults in depth in
Section~\ref{sec:protocol}.
These failures can be intentional (to
gain application level benefits) or unintentional (due
to software bugs or external attacks); \system does not
distinguish between the two.

%\begin{itemize}
% \textbf{{1) Datastore Layer}}: A faulty server can
%     corrupt the data shard it stores by intentionally changing the content. Corruption
%     may occur during the execution or commitment or post
%     commitment of a transaction. 
    
%     \textbf{{2) Transaction Execution Layer (TEL)}}: This layer can return wrong timestamps or inconsistent 
%     values to the client or buffer incorrect writes. To minimize 
%     the possible faults in this layer, 
%     it is intentionally kept light-weight.
    
%   \textbf{{3) Transaction Commitment Layer (TCL)}}: A
%     faulty commit layer can: i) break the transaction 
%     correctness by locally deciding to commit
%     non-serializable transactions, ii)
%     violate transaction atomicity by maliciously convincing some
%     servers to \textsf{commit} and others to \textsf{abort}
%     a transaction. 
    
%     \textbf{{4) Tamper-Proof Log}}: A faulty server
%     can modify its local copy of the transaction
%     log and either omit or reorder the transaction execution 
%     order. Any subset of servers (but not all) can
%     collude and attempt to produce a false log.
    
% %\end{itemize}
A malicious client in \system
can send arbitrary messages or semantically 
incorrect transactions
to a database server
but later blame the server for updating the database
inconsistently. To circumvent this, the servers 
store all digitally signed, unforgeable messages
exchanged with the client. This message log
serves as a proof against a falsified blame or
when a client's transaction sends the database
to a semantically inconsistent state.

\subsection{Auditing \system}
\label{sub:audit}
Auditability
has played a key role in building dependable
distributed systems~\cite{yumerefendi2004trust, 
yumerefendi2005role, haeberlen2007peerreview}.
\system provides 
auditablility: the application layer or an external 
auditor can audit individual servers 
with an intent to either detect failures or verify 
correct behavior. 

\system guarantees that any failure,
as discussed in
Section~\ref{sec:failure_model}, will 
be detected in an offline audit.
\system focuses on failure detection rather than prevention;
detection includes identifying (i) the \textbf{precise point} 
in transaction history where an anomaly
occurred, and (ii) the exact misbehaving server(s) that
is irrefutably
linked to a failure.

The auditor is considered to be a powerful
external entity and during each audit:

(i) The auditor gathers the tamper-proof
logs from all the servers before the auditing process.

(ii) Given that at least one server is correct, 
from the set of logs collected from all servers,
the auditor identifies the \textit{correct} and
\textit{complete} log (how is explained in detail in
Section~\ref{sub:log}). The auditor uses this log
to audit the servers.

Optimizations such as checkpointing
\cite{koo1987checkpointing} can be used to
minimize the log storage space at each
server; these optimizations are orthogonal
and hence not discussed further.
If the audit uncovers any malicious activity, a 
practical solution can be to penalize the misbehaving
server in legal, monetary, or other forms specific to the 
application. This discourages a server from
acting maliciously.
\vspace{-0.5em}
\section{Fides}
\label{sec:protocol}

\begin{figure}[t!]
\centering
\includegraphics[scale=0.42]{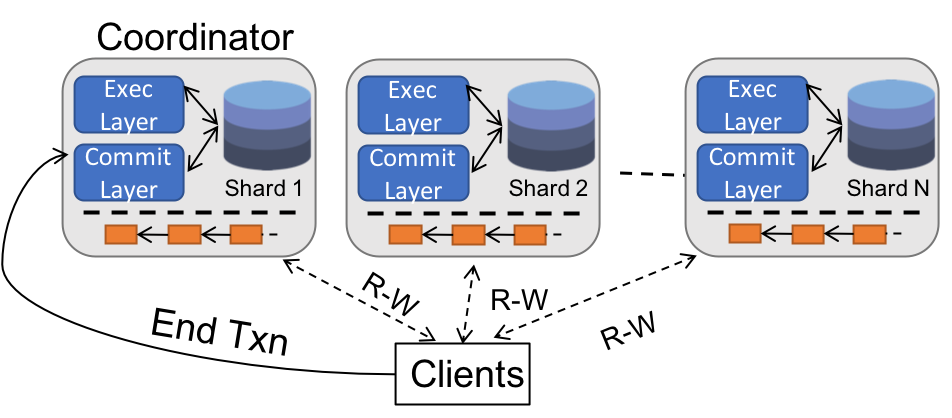}
\caption{Client interactions in \system}
\label{fig:arch}
\vspace{-1em}
\end{figure}

In this section we present \system: an
\textit{auditable} data management system built on  
untrusted infrastructure.
The basic idea is to integrate crypotographic techniques 
such as digital signatures (public and private key encryption), 
collective signing, and Merkle Hash Trees (MHT) with
the basic transaction execution in database systems. This
integration results in \emph{verifiable} 
transaction executions in an environment where the database 
servers cannot be trusted.

\subsection{Overview}
Figure~\ref{fig:arch} illustrates the overall design of
\system. The clients read and write 
relevant data by directly interacting with the appropriate
database partition server (this can be accomplished by linking
the client application with a run-time library that provides a
lookup and directory service for the database partitions). 
The architecture intentionally avoids the layer of 
front-end database servers (e.g., Transaction Managers) 
to coordinate the execution of transaction 
reads and writes as these
front-end servers may themselves be
vulnerable and exhibit malicious behavior by relaying
incorrect reads/writes. Hence, in \system all data-accesses
are managed directly between the client and the relevant
database server.
%No other intermediary is involved in handling reads 
% and writes from the client.

Since data-accesses are handled with minimal synchronization 
among concurrent activities, the burden of 
ensuring the correct execution of transactions occurs when a
transaction is \emph{terminated}. We use a simplified setup 
where one \textit{designated} server acts as the transaction 
\textit{coordinator} responsible for terminating all
transactions. The coordinator is also an
untrusted database server that has additional 
responsibilities only during the termination phase.

When a client application decides to terminate its transaction, 
it sends the termination request to the designated coordinator; 
all other database servers act as cohorts during the
termination phase. For ease of exposition, we first
present a termination protocol executed
globally involving \textbf{\textit{all}} database 
servers, irrespective
of the shards accessed in that transaction. The
global execution implies transactions are terminated
\textit{sequentially}. Later we relax this requirement
and allow different coordinators for
concurrent transactions. 

% A main constraint
% that has to be maintained during this coordination is \textit{confidentiality} i.e., at no
% point does a database server $S$ observe the data-values of data 
% stored at another database server $S'$.  
\begin{figure}[t!]
\centering
 \includegraphics[scale=0.36]{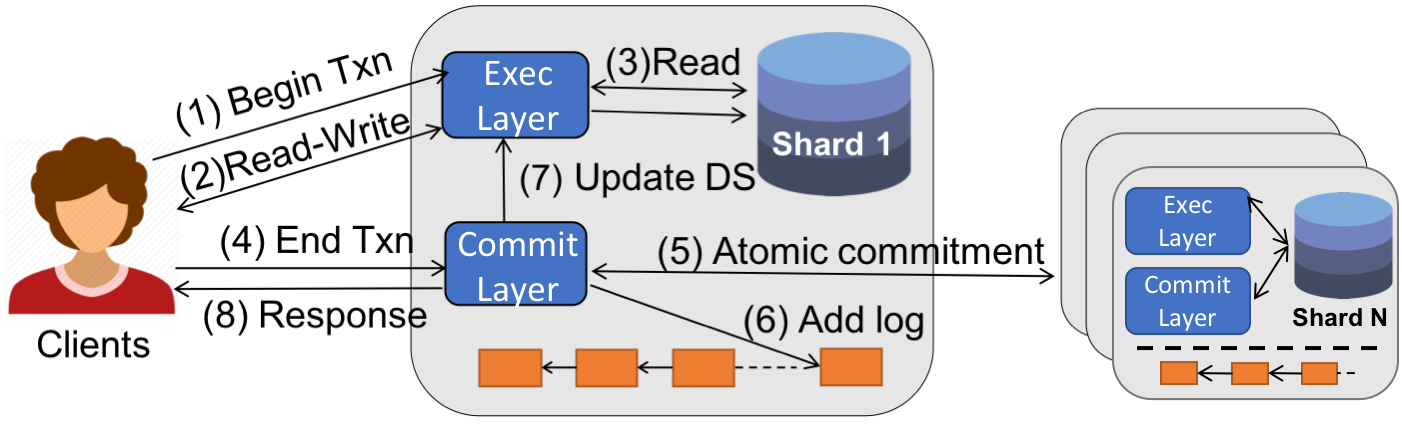}
\caption{Transaction life-cycle in \system}
\label{fig:deployment}
\vspace{-1em}
\end{figure}

The following is an overview of the 
client-server interaction: a typical
life-cycle of a transaction as depicted in
Figure~\ref{fig:deployment}.

\noindent\textbf{1. Begin transaction}: A client starts
accessing the data by first sending a \textit{Begin 
Transaction} request to all the database
servers storing items read or written by the transaction.

\noindent\textbf{2. Read-write request}: The client then sends 
requests to each server indicating the
data items to be read and written.

\noindent\textbf{3. Read-write response}: The transaction execution
layer responds to a read request by fetching
the data from the datastore and relaying it to the client.
The write requests are buffered.

\noindent\textbf{4. End Transaction}: After completing 
data access, the client sends \textit{End Transaction}
to the coordinator which coordinates the commitment
to ensure transaction correctness
(i.e., serializability) and transaction atomicity (i.e., 
all-or-nothing property).
% and for
% potential malicious behavior of agents involved in the transaction
% execution. 
%This protocol is presented in Section~\ref{sub:tcl}.

\noindent\textbf{5. Atomic commitment}: The coordinator
and the cohorts collectively execute the atomic commit 
protocol -- \protocol -- and decide either to \textsf{commit} 
or \textsf{abort}
the transaction. The commitment produces a block
(i.e., an entry in the log)
containing the transaction details. If the 
decision is \textsf{commit},
then the next two steps are performed.

\noindent\textbf{6. Add log}: All servers append,
to their local copy of the log, the same block in a 
consistent order, thus creating a globally 
replicated log.

\noindent\textbf{7. Update datastore}: The datastore
is updated based on the buffered
writes, if any, along with updating the
timestamps $r_{ts}$ and $w_{ts}$ of the data items
accessed in the transaction.

\noindent\textbf{8. Response}: The coordinator 
responds to the client informing whether the transaction
was committed or aborted.

The log, stored as a linked-list of blocks,
encompasses the transaction details essential
for auditing. It is vital to understand the
structure of each block
before delving deeper into the transaction execution
details. Every block
stores the information shown in Table~\ref{tab:block}.
Although a block can store multiple transactions,
for ease of explanation, \textit{we assume that
only one transaction is stored per block}. 

\begin{table}[]
\centering
\begin{tabular}{ c | c }  
\rowcolor{LightCyan}
 \textit{\textbf{key}} & \textit{\textbf{description}} \\
\hline
\hline
 TxnId & commit timestamp of txn \\ 
 \rowcolor{Gray}
 R set & list of $\langle{id: value, r_{ts}, w_{ts}}\rangle$ \\  
 W set & list of $\langle{id: new\_val, old\_val, r_{ts}, w_{ts}}\rangle$ \\ 
 \rowcolor{Gray}
 $\small\sum roots$ & MHT roots of shards  \\
 $decision$ & \textsf{commit} or \textsf{abort}\\
 \rowcolor{Gray}
 $h$ & hash of previous block \\
 $co\mbox{-}sign$ & a collective signature of participants 
\end{tabular}
\caption{Details stored in each block}
\label{tab:block}
\end{table}

As indicated in Table~\ref{tab:block}, each transaction
is identified by its commit timestamp, assigned by the
client that executed this transaction. Any timestamp that 
supports total ordering can be used by the client -- e.g., 
a Lamport clock with $\langle client\_id:client\_time\rangle$ --
as long as all clients use the same timestamp generating
mechanism.

A block contains the transaction read and write sets
consisting of three vital pieces
of information: 1) the data-item identifiers that are read/written, 
2) the values of items read and the new
values written; the $old\_val$ in the write set is
populated only for blind writes, and 3) 
the latest read $r_{ts}$ and write $w_{ts}$ timestamps
of those data items at the time of access (read or
write).

The blocks also contain: the
Merkle Hash Tree roots of the shards involved in the transaction
(explained more in Section~\ref{sub:ccl});
the \textsf{commit} or \textsf{abort} 
transaction decision; the hash of the previous block
forming a chain of blocks linked by their
hashes; and finally, a collective signature of all the servers 
(how and why are explained in Section~\ref{sub:tcl}).

The following subsections elaborate 
on the functionalities
of a database server in a transaction life cycle.
For each functionality, we first explain
the correct behavior followed by the techniques
to detect malicious faults. 
 
\subsection{Transaction Execution}
\label{sub:ccl}

This section describes the correct mechanism for
executing transactions (reads and writes) and 
discusses techniques to detect deviations from the
expected behavior. 
% The clients are aware of the server-shard mapping (which
% database server stores what shard); and during
% the execution phase, a client interacts only with those
% servers whose data are being read/written in the transaction.
% The execution layer is a light-weight layer
% with a sole responsibility of responding to the read/write 
% requests from different clients.

\begin{figure}[h!]
\centering
\includegraphics[scale=0.4]{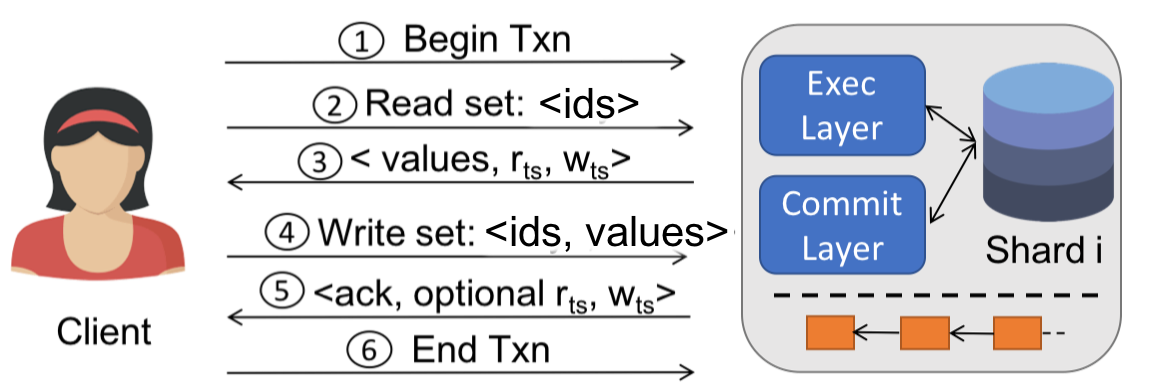}
\caption{Transaction execution in \system}
\label{fig:txn_exec}
\vspace{-1em}
\end{figure}

\subsubsection{Correct Behavior}
Figure~\ref{fig:txn_exec} depicts the 
client-server interactions during transaction execution. 
With regard to transaction execution, 
a correct database server is responsible for the
following actions: (i) return the values and timestamps
of data-items specified in the read requests, and 
(ii) buffer the values of data-items
updated in the transaction and if the transaction 
successfully commits, update the datastore based on
the buffered writes. We explain how a correct server
achieves these actions.
% The clients interact with
% the data via transactions comprising of read and write
% operations. 

\textbf{\textit{Reads and Writes}}: 
A client
sends a \textit{begin transaction} message 
to all the database
servers storing the items read or written by the transaction.
The client then sends a \textit{Read} request consisting
of the data-item ids to the 
respective servers. For example, if a transaction
reads data item $x$ from server $S1$ and item $y$
from server $S2$, the client sends \textit{Read(x)} to $S1$
and $Read(y)$ to $S2$.
The servers respond with
the data values
\textbf{\textit{along with}} the associated read
$r_{ts}$ and write $w_{ts}$ timestamps. 

The client then sends the \textit{Write}
message with the data-item ids and their 
updated values to the respective servers. For example, 
if a transaction
writes data item $x$ in server $S1$ with value $5$
and item $y$
in server $S2$ with value $10$,
the client sends \textit{Write(x,5)} to $S1$
and \textit{Write(y,10)} to $S2$.
The servers buffer these updates and respond with an
acknowledgement. To support blind
writes, the acknowledgement includes 
the old values and associated 
timestamps of the data-items that are being written
but were not read before.

After completing 
the data accesses, the client sends the \textit{end 
transaction} request -- sent
only to the designated coordinator --
consisting of the read and the write set: a list of
data item ids, the corresponding timestamps $r_{ts}$ and  $w_{ts}$ returned by the servers,
and the values read and the new values written.
The coordinator then executes \protocol among all
the servers to
terminate (commit or abort) the transaction
(explained in detail in Section~\ref{sub:tcl}).
If all the involved servers decide to commit
the transaction, each involved server constructs a
Merkle Hash Tree (MHT) (Section~\ref{sub:mht}) 
of its data shard with all the data items -- with
updated values -- as the leaves of the tree 
and with the root node $root_{mht}$.
The read and write sets and MHT roots
become part of the block
in the log once the transaction is committed.

\textbf{\textit{Updating the datastore}}:
If the transaction commits, 
the servers involved in the transaction
update the data values in their
datastores based on the
buffered writes. The servers also update the
read and write timestamps of the data items
accessed in the transaction
to the transaction's commit timestamp.

The data can be single-versioned or 
multi-versioned.
For multi-versioned data, when a transaction 
commits, a correct server additionally creates
a new version of the data items accessed in the
transaction \textit{while maintaining the older versions}.
Although an application using \system can choose between
single-versioned
or multi-versioned data, multi-versioned data
can provide \textit{\textbf{recoverability}}.
If a failure occurs,
the data can be reset to the last sanitized 
version and the application can resume execution from 
there.

\subsubsection{Detecting Malicious Behavior}
\label{sub:ccl_verify}
With regard to transaction execution, 
a server may misbehave by: (i) returning inconsistent 
values of data-items specified in the read
requests; and (ii) buffering incorrect values  of 
data-items updated in the transaction
or updating the datastore incorrectly. 

\textbf{\textit{(i) Incorrect Reads}}:
All faults in \system are detected by an 
auditor during an audit. As mentioned in
Section~\ref{sub:audit}, during an audit,
the auditor collects
the log from all servers and constructs the correct
and complete log.

To detect an incorrect read value returned by a
malicious server, the auditor must know the expected
value of the data-item. The read and write sets
in each log entry contains the information on 
the updated value of a written item and the
read value of a read item.
Note that in our simplifying assumption (which will
be relaxed later), each block contains
\textit{only one} transaction and the transactions
are committed sequentially with the log reflecting
this sequential order. By traversing the log, at each
entry, the auditor knows the most recent values
of a given data item. We leverage
this to identify incorrectly returned values. 

\vspace{0.5em}
\noindent\textbf{\textit{Lemma 1}}: The auditor
detects an incorrect value returned for a data
item by a malicious server.

\textbf{\textit{Proof}}: Consider a transaction $T_i$ that
committed at timestamp $ts_i$ and stored in the log
at block $b_i$. Assume transaction $T_i$ read
an item $x$ and updated it. Let $b_j$ be
the first block after $b_i$ to
access the same data item $x$ -- where $j>i$,
indicating that transaction $T_j$ in $b_j$ committed
\textit{\textbf{after}}
the transaction $T_i$ in $b_i$. 
The read value of $x$ in $b_j$ must reflect
the value written in $b_i$; if the values
differ, an anamoly is detected.\hfill$\Box$

\vspace{0.5em}
\textbf{\textit{(ii) Incorrect Writes}}: The effect
of incorrectly buffering a write or incorrectly
updating the datastore is the same: the datastore
ends up in an inconsistent state.
The definition of incorrect datastore depends
on the type of data: for single versioned data, 
the latest state of data (data values and timestamps)
in the datastore
is incorrect; for multi-versioned data, 
one or more versions of the data are incorrect. We
discuss techniques to detect incorrect datastore
for both types of data.

To detect an inconsistent datastore, we use the data 
authentication technique proposed by
Merkle~\cite{merkle1989certified} discussed in
Section~\ref{sub:mht}. To use this technique,
the auditor requires the read and written values
in each transaction and the resultant
Merkle Hash Tree (MHT) root -- all pieces of
information
stored within each block. 

\textit{\textbf{Multi-versioned data:}}
For multi-versioned data, the audit 
policy can involve auditing a single version chosen 
arbitrarily or exhaustively auditing all versions
starting from either the first version
(block 0) or the latest version. We explain 
auditing a single version, which can easily be extended to
exhaustively auditing all versions.

Let $T_i$ be a transaction committed at timestamp
$ts$ that read and wrote data item $x$ 
stored in server $S_k$.
Assume the auditor audits server $S_k$ at version $ts$. Once the auditor
notifies the server about the audit,
the
server constructs the Merkle Hash Tree with the
data at version $ts$
as the leaves; $S_k$ then shares the 
\textit{Verification Object} \vo -- consisting of all
the sibling nodes along the path from the data $x$
to the root -- with the auditor.

The log entry corresponding to transaction $T_i$ 
stores the value read for item $x$ and the
new value written.
The auditor uses (i) the \vo sent by $S_k$, and (ii) the 
hash of $x$'s value stored in the write set of the log,
to compute the
expected MHT root for the data in $S_k$ (discussed in
Section~\ref{sub:mht}). The auditor
then compares the computed root with the one stored in 
the log. A mismatch indicates that the data at version
$ts$ is incorrect.

\textit{\textbf{Single-versioned data:}}
For single versioned data, the correctness is only with
respect to the latest state of the data. Hence,
rather than using an arbitrary block to obtain
the MHT root of server $S_k$, the auditor uses the latest
block in the log that accessed
the data in $S_k$ to obtain the latest MHT root. 
The other steps are similar to multi-versioned data: 
the auditor fetches the \vo based on the latest state 
of $S_k$ and recomputes
the MHT root to compare the root stored in the 
log. 

\vspace{0.5em}
\noindent\textbf{\textit{Lemma 2}}: The auditor
detects an inconsistent datastore.
For multi-versioned data, the auditor
detects the precise version at which
the datastore became inconsistent.

\textit{\textbf{Proof}}:
Detection is guaranteed since Merkle Hash Trees (MHT) use
collision-free
hash functions (i.e., $h(x) \neq h(y)$ where $x \neq y$), and a malicious
server cannot update a data value such that the MHT root
stored in the block
matches the root computed by the auditor using the
verification object sent by the server. 
% For
% single versioned datastores, the latest block in the log
% to access the data is used to extract the MHT
% root. 
For multi-versioned datastores, the auditor identifies
the precise version at which data corruption occurred 
by systematically authenticating all blocks in the log
until a 
version with mismatching MHT roots is detected.
\hfill$\Box$
\subsection{Transaction Commitment}
\label{sub:tcl}

This section describes how transactions are
terminated in \system and presents a novel 
distributed atomic commitment protocol --
\textbf{\textit{TrustFree 
Commit}} (\protocol) -- that handles malicious failures.
This section also discusses techniques to detect
failures if a server deviates from the expected behavior.
With regard to transaction commitment, a correct database
server is responsible for the following actions:
(i) Ensure transaction isolation (i.e., strict 
serializability);
(ii) Ensure atomicity -- either all servers
commit the transaction or no servers commit the transaction;
and (iii) Ensure \textit{verifiable} atomicity.

\subsubsection{Correct Behavior}
\textbf{\textit{Transaction Isolation}}: 
Transaction isolation
determines how the impact of one transaction is
perceived by the other transactions. In \system,
even though multiple transactions can execute concurrently, \system provides
serializable executions in which concurrent transactions
seem to execute in sequence. To do so, servers in \system
abort a transaction if it cannot be serialized 
with already committed transactions in the log. The
read $r_{ts}$ and write $w_{ts}$ timestamps associated
with each data item is used to detect non-serializable
transactions. The latest timestamps can be obtained
from either the datastore or the transaction log. 
Similar to timestamp based optimistic concurrency
control mechanism, at commit time, a server
checks if the data accessed in the terminating
transaction has been updated since they were read.
If yes, the server chooses to abort the transaction.
% For example, consider two concurrent
% transactions $T_i$ and $T_j$. $T_i$ read data item $x$ at 
% $r_{ts}$=\textit{ts-100}; but later,
% $T_j$ updates $x$ and commits at timestamp
% \textit{ts-120}. If  $T_i$ tries to commit \textit{after} $T_j$, the two transactions become
% non-serializable, and hence
% $T_i$ must be aborted.

\textit{\textbf{Atomicity and Verifiablity:}} 
Consider a traditional atomic commit protocol
that provides atomicity: Two Phase Commit 
(2PC)~\cite{gray1978notes}. 
2PC guarantees atomicity provided
servers are benign and trustworthy.
It is a centralized protocol where one server acts as
a coordinator and the others act as cohorts. To
terminate a transaction, the coordinator collects 
\textsf{commit} or \textsf{abort} votes from all cohorts,
and decides to \textsf{commit}
the transaction \textit{only if} all the cohorts choose
to commit, and otherwise decides to \textsf{abort}. The
decision is then asynchronously sent
to the client and the cohorts.
2PC is sufficient to ensure atomicity if servers
are trustworthy; but in untrusted environments,
2PC is inadequate as
a cohort or the coordinator may maliciously lie about
the decision. We need to develop an atomic commitment 
protocol that can overcome such malicious behaviour.

\begin{figure*}[ht!]
\centering
\includegraphics[scale=0.5]{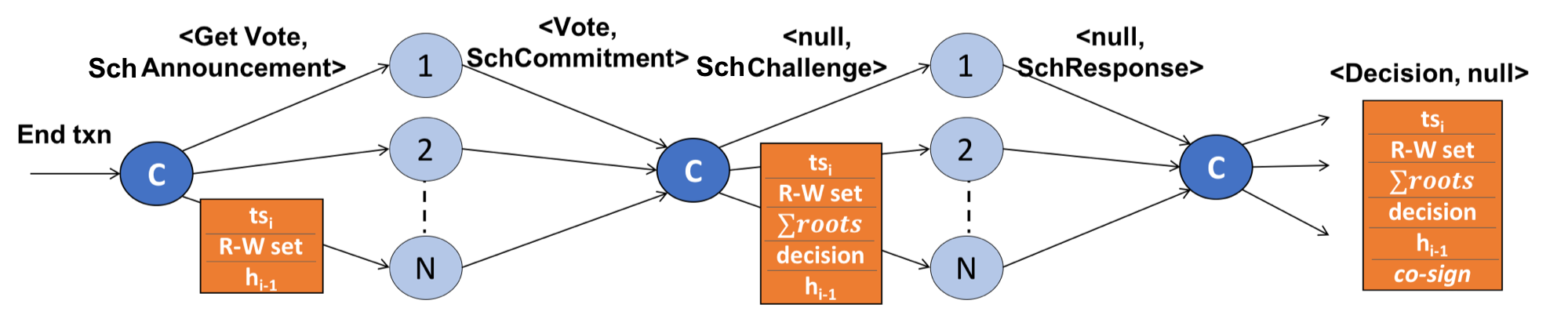}
\caption{Different phases and block generation progress made in each phase of \protocol}
\label{fig:txn_commit}
\vspace{-0.5em}
\end{figure*}
To make 2PC trust-free, we combine 2PC with a
multi-signature scheme,
Collective Signing or CoSi (Section~\ref{sub:cosi}): a
two-round protocol
where a set of processes collectively sign a given record 
using their private keys and random secrets.
CoSi guarantees that a record (or in our case block) 
produced by a leader (or
coordinator) is validated and signed by all the
witnesses (or cohorts) and that \textit{if any of the involved
processes lied in any of the phases, the resulting 
signature will be incorrect}. A signature is bound to
a single record; any process with the 
public keys of all the processes can verify whether 
the signature is valid and \textit{corresponds} to that 
record.

We propose a novel approach of integrating 2PC with CoSi
to achieve the atomicity properties
of 2PC \textit{and} the verifiable properties of CoSi.
The basic idea is that the coordinator, similar to 2PC,
collects \textsf{commit} or \textsf{abort} 
votes from the cohorts, forms a decision, and
encapsulates the transaction details including
the decision in a block. The coordinator then
sends the block to be verified and collectively 
signed by the cohorts. An incorrect block (either with 
inaccurate transaction details or wrong decision) 
produced by a malicious
coordinator will not be accepted by correct servers, thus
resulting in an invalid signature that can be easily 
verified by an auditor.\\

A successful round of \protocol produces 
a block to be appended to the log \textit{in a consistent
order} by all servers.
For ease of exposition, this section presents
\protocol with two main assumptions: (i) the transactions
are {committed sequentially} to avoid forks in the log;
and (ii) {all} servers participate 
in transaction termination -- even the servers 
that did not partake in transaction execution --
to have identical block order in their logs.
In Section~\ref{sub:scaling} we relax these assumptions
and discuss various techniques to scale \protocol.

Recall from Table~\ref{tab:block} all the details
stored in each block. Once a block is cosigned and
logged by all servers, it is immutable;
hence, all the details must be filled in during different
phases of \protocol. However, to ensure atomicity and
verifiability of \protocol, we only need the transaction
id, its decision, and the co-sign.
Other details such as the \textit{Read} and
\textit{Write} sets, Merkle Tree roots, and hashes
are %used only by the auditor as they are 
necessary to detect other failures including
isolation violation and data corruption.

\vspace{0.5em}
\noindent\textbf{\textit{The protocol:\\}}
A client, $\mathcal{A}$, upon finishing transaction 
execution, sends a
signed $\mu = \big\langle
end\_transaction(T_{id}, ts_i, R\mbox{ }set\mbox{-}W set)\big\rangle_{\sigma\mathcal{A}}$ request
to the coordinator, where $T_{id}$ is a unique 
transaction id and $ts_i$ is a client-assigned
commit timestamp of the transaction.
The request also includes $R\mbox{ }set\mbox{-}W set$:
the read and write sets
consisting of data item ids, values 
read and new values written, $r_{ts}$, and $w_{ts}$.
The servers 
ignore any \textit{end transaction} request with
a timestamp lower than the latest committed timestamp.

\protocol is a 3-round protocol involving
5 phases of communication as shown in
Figure~\ref{fig:txn_commit}. Since \protocol
merges 2PC with CoSi, we indicate each phase
by a mapping of $<$2PC phase, CoSi phase$>$.
Figure~\ref{fig:txn_commit} shows the phases
as well as the progress made in constructing
the block at each phase. 
The phases of \protocol are:

\textbf{1) $<$GetVote, SchAnnouncement$>$}: Upon receiving the
    $\mu =
\big\langle
end\_$ $transaction(T_{i}, ts_i, R\mbox{ }set\mbox{-}W set)\big\rangle_{\sigma\mathcal{A}}$ request from the client,
    to commit transaction $T_i$, the coordinator $\mathcal{C}$ 
    prepares a partially filled block,
    $b_i=[ts_i, R set\mbox{ - }W set, h_{i-1}]$, containing the commit
    timestamp, read and write sets, and hash of the previous block.
    $\mathcal{C}$ then 
    encapsulates the signed client request $\mu$ and sends 
    the $\big\langle get\_vote(b_i, \mu)\big\rangle_
    {\sigma  C}$ message to all the cohorts. 

\vspace{0.5em}
\textbf{2) $<$Vote, SchCommitment$>$}:  Every cohort  $\mathcal{H}$ 
    verifies 
    both the get\_vote message and the encapsulated client 
    request, and 
    computes the Schnorr-commitment ($x_{sch}$) for CoSi.
    Then, \textit{only the cohorts that are part of the transaction},
    perform the following actions. A cohort 
    involved in the transaction locally decides whether to
    \textsf{commit}
    or \textsf{abort} the transaction. If the cohort
    locally decides to
    \textsf{commit}, then it constructs a Merkle Hash
    Tree (MHT) (Section~\ref{sub:mht}) of its shard with all the
    data items as leaves of the MHT and with the root node 
    $root_{mht}$. The MHT reflects all the 
    updates in $T_i$ assuming that $T_i$ be committed; since MHT
    computation is done in 
    memory, the datastore is unaffected if $T_i$ 
    eventually aborts. (The MHT root is 
    required for datastore authentication, as 
    explained in Section~\ref{sub:ccl_verify}.)
    The involved cohorts
    then send $\big\langle vote(decision, root_{mht}, x_{sch})\big\rangle_
    {\sigma\mathcal{H}}$ whereas the cohorts not part of the
    transaction send $\big\langle vote(x_{sch})\big\rangle_
    {\sigma\mathcal{H}}$
    to the coordinator. As the coordinator is also
    involved in co-signing, it produces the
    appropriate vote message.
\vspace{0.5em}

\textbf{3) $<$null, SchChallenge$>$}: In this phase, the coordinator
    $\mathcal{C}$ collects all the cohort responses and
    checks if any cohort (or itself) involved in the transaction 
    decided to abort.
    If none, it chooses \textsf{commit}, otherwise \textsf{abort}.
    It then aggregates all the MHT roots of the involved cohorts 
    ($roots = \sum root_{mht}$),
    and fills the roots field in the block $b_i$ along with the
    decision field. If any involved cohorts chose 
    \textsf{abort}, the respective roots will be missing in the
    block. Finally, the coordinator aggregates the 
    Schnorr-commitments $X_{sch}=\sum x_{sch}$ from all the
    servers and computes 
    the Schnorr-challenge by concatenating and hashing $X_{sch}$ with $b_i$ \textit{i.e.,}
    $ch = h(X_{sch} || b_i)$. 
    The coordinator then sends  $\big\langle challenge(
    ch, X_{sch}, b_i)\big\rangle_{\sigma C}$ to all cohorts.
\vspace{0.5em}

\textbf{4) $<$null, SchResponse$>$}: In this phase, every cohort,
    $\mathcal{H}$, checks if the decision within the block $b_i$
    is \textsf{abort}, and if so, $b_i$ should
    have some missing roots; if the decision is \textsf{commit}, 
    $b_i$ should have all the roots from the involved servers.
    Every involved cohort that sent the MHT root in 
    the \textit{vote} phase verifies if its corresponding root
    in the block is the same as the one it sent.
    Cohorts also verify whether a potentially
    malicious coordinator 
    computed the challenge, $ch$, correctly by hashing the
    concatenated $X_{sch}$ and $b_i$, both of
    which were sent in the challenge message. A cohort
    then computes the Schnorr-response
    $r_i$ using its secret key and the challenge $ch$, and sends
    $\big\langle response(r_i)\big\rangle_{\sigma\mathcal{H}}$
    to the coordinator.

\textbf{5) $<$Decision, null$>$}:
    The coordinator collects all the Schnorr-responses
    and aggregates them, $R_{sch}=\sum r_{sch}$, 
    to form the collective signature represented by 
    \textbf{$\langle ch, R_{sch} \rangle$}. Intuitively,
    the challenge $ch$ is computed using the block; and the
    Schnorr-response $R_{sch}$ requires the private
    keys of the servers, thus the signature binds
    the block with the public keys of the servers. The coordinator
    then updates the
    \emph{co-sign} field in the block and sends the 
    finalized block to the
    client and the cohorts. If the decision is commit,
    all servers
    append block $b_i$ to their log and update their respective datastores.\\

The client, with the public keys of all the servers, verifies the
co-sign before accepting the decision -- even an aborted 
transaction must be signed by all the servers.
If the verification fails, the client detects an
anomaly and 
triggers an audit, which may halt the progress in the 
system. 

% \protocol requires \textit{all} servers to participate
% in the transaction termination. 
\protocol, similar to 2PC,
can be blocking if either the coordinator or any cohort 
fails (crash or malicious). \protocol
can be made non-blocking by adding
another phase that makes the chosen value available,
as in the case of Three Phase Commit~\cite{skeen1981nonblocking};
we leave this
extension for future work.
\subsubsection{Detecting Malicious Behavior} 
A correct execution of \protocol ensures serializable 
transaction isolation, atomicity, and verifiable
commitment. However, a malicious server can (i) violate 
the isolation guarantees by committing non-serializable
transactions; (ii) a malicious coordinator can break
atomicity by convincing some servers to commit a
transactions and others to abort; or (iii) a server can
send wrong cryptographic values during co-signing
to violate verifiability.

\vspace{0.5em}
\noindent\textbf{\textit{Lemma 3}}: The auditor 
detects serializablity violation.

\textbf{\textit{Proof}}:
Transaction execution is based on executing read and write
operations in the timestamp order. The transactions
are ordered based on the timestamps, which are monotonically
increasing. If a transaction has done a conflicting access
inconsistent with the timestamp order, it leads to one of the
following conflicts:
1) RW-conflict: a transaction with a smaller timestamp read a data-item with a larger timestamp;
2) WW-conflict: a transaction with a smaller timestamp wrote a data-item that was already updated with a larger timestamp;
3) WR-conflict: a transaction with a smaller timestamp wrote a data-item after it was read by a transaction with a larger timestamp.
For each transaction audited, the auditor verifies if any of
the above violations exist, and if so, the auditor detects
the server responsible for the violation to be misbehaving.
This is equivalent to verifying that no cycle exists in the
Serialization Graph of the transactions being audited.\hfill$\Box$

\vspace{0.5em}
\noindent\textbf{\textit{Lemma 4}}: 
The auditor or a correct server detects incorrect cryptographic values for CoSi sent by a malicious server -- which hampers verifiablity of \protocol.

\textbf{\textit{Proof}}:
If any server sends an incorrect cryptographic value used
for co-signing, this results in an invalid signature, and
the original work CoSi~\cite{syta2016keeping} 
guarantees identifying the precise server that computed
the crytographic values incorrectly. Since \protocol
incorporates CoSi, it inherits this guarantee
from CoSi. Intuitively, in the \textit{schResponse}
phase, the coordinator can identify if the
signature is invalid, in which case, it can check
partial signatures produced by excluding one 
server at time and detect the precise server
without which the signature is valid. The
coordinator is incentivised to perform this
rigorous check because if the signature is
invalid, the auditor suspects the coordinator for
producing an incorrect block. We refer to the original
work~\cite{syta2016keeping} that 
discusses the proof in depth.\hfill$\Box$

\vspace{0.5em}
\noindent\textbf{\textit{Lemma 5}}: 
The auditor or a correct server
detect atomicity violation of \protocol.
% If a malicious coordinator
% violates the atomicity of \protocol, a correct server or
% an auditor will
% detect it.

\begin{figure}[t!]
\centering
\includegraphics[scale=0.38]{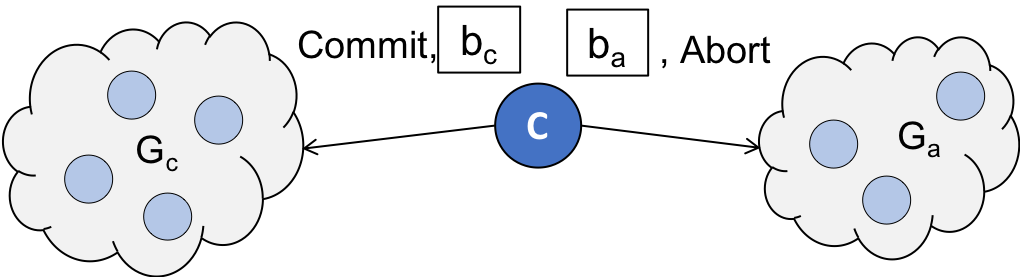}
\caption{Atomicity violation of \protocol}
\label{fig:commit_violation}
\vspace{-1em}
\end{figure}

\textbf{\textit{Proof}}:
Recall that the coordinator $\mathcal{C}$ collects 
votes in phase two of \protocol, forms the decision, and
sends the partial block containing the decision in 
the \textit{challenge} message.
Consider Figure~\ref{fig:commit_violation} where a
malicious coordinator sends 
block $b_c$ with \textsf{commit} decision to group $G_c$
and block $b_a$ with \textsf{abort} decision to 
group $G_a$.
More precisely, the coordinator sends $\big\langle challenge(
ch, X_{sch}, b_c)\big\rangle_{\sigma C}$ to $G_c$ 
($X_{sch}$ is the aggregated Schnorr-commits) and
$\big\langle challenge(ch, X_{sch}, b_a)\big\rangle_
{\sigma C}$ to $G_a$. Since the decision is part of
the block, the two blocks $b_c$ and $b_a$
have to be different if the coordinator violates
atomicity. But with respect to the challenge $ch$,
there are two possibilities, both producing invalid
signatures:

$\bullet$ \textit{Case 1}: Coordinator sends the same challenge $ch$
computed using block $b_c$ (or $b_a$) to both groups.

Any correct server in the group $G_a$ 
will recompute the challenge using the block
it received, $b_a$, and immediately
recognize that the challenge sent by the coordinator
does not correspond to the block $b_a$.
(Alternatively, if the coordinator used $b_a$ to 
compute the challenge $ch$, then servers
in $G_c$ will detect the anomaly.) Even if the 
servers in one group, say $G_a$, collude with the
coordinator and do not expose the anomaly, the
challenge $ch$ corresponds only to block $b_c$.
The auditor, while auditing a server in group 
$G_a$, detects that the co-sign in block $b_a$
is invalid as it does not correspond to that block.

\vspace{0.5em}
$\bullet$ \textit{Case 2}: Coordinator sends the challenge $ch$
computed using block $b_c$ to group $G_c$ and the challenge
$ch^\prime$ computed using block $b_a$ to group $G_a$.

In the final step of \protocol,
the servers in group $G_c$ will use $ch$ to compute 
the Schnorr-response, whereas the servers in group $G_a$
will use $ch^\prime$ to compute the Schnorr-response. Given
that the final collective signature can be
tied only to a single block, the co-sign
does not correspond to either $b_c$ or $b_a$,
hence producing a wrong signature.\hfill$\Box$ 

\vspace{0.5em}
The coordinator or a cohort
\textbf{can never force} all
servers to commit if at least one server decides to abort
a transaction. For committed transaction,
the transaction block must contain MHT roots from
all the involved servers; for aborted transactions, the
block should have at least one MHT root missing.
Assume a server $S_b$ chooses abort and hence, does
not send its MHT root. If the coordinator
produces a fake root for server $S_b$, the server
will detect it
in the \textit{schResponse} phase.
And in case
server $S_b$ colludes with the coordinator by either
not exposing the fake root or by producing a
fake root itself,
the datastore verification (discussed in Section
\ref{sub:ccl_verify}), which uses MHT roots,
will fail for server $S_b$. An involved server
(coordinator or cohort)
can only force an abort on all servers by
choosing to abort the transaction, which is tolerable 
as
the decision will be consistent across all servers
and will not violate the atomicity of \protocol.

\subsection{Transaction Logging}
\label{sub:log}

The transaction log in \system is a tamper-proof,
globally replicated log. When a transaction commits
after a successful round of \protocol, all servers
append the newly produced block to their logs. 

\textbf{Detecting Malicious Behavior}: One or more
faulty servers can collude (but not all at once) to
(i) tamper an arbitrary block, (ii) reorder the
blocks, or (iii) omit the tail of the log (last few 
blocks). The auditor collects logs from all the 
servers and uses the collective signature 
stored in each block to detect
an incorrect log.

\vspace{0.5em}
\noindent\textbf{\textit{Lemma 6}}: Given a set
of logs collected from all servers, the auditor detects
all incorrect logs -- logs with arbitrary 
blocks that are modified or logs with reordered blocks.

\textit{\textbf{Proof}}:  The collective signature in
each block prevents a malicious server from manipulating 
that block once it is appended to the log. The signature 
is tied specifically to one block and if the contents
of the block
are manipulated, the signature verification will fail.
One or more malicious servers cannot tamper with an
arbitrary block successfully without the
cooperation of \textit{all} the servers.
And since the hash of the previous block is part of a log
entry, unless \textit{all} the servers collude, the
blocks cannot be successfully re-ordered.
% The hash functions used for chaining the blocks in
% the log prevents a server from re-ordering the blocks. 
% Due to collision 
% free property of hash functions ($h(b_i) \neq h(b_j)$),
% two blocks $b_i$
% and $b_j$ cannot be successfully reordered.
\hfill$\Box$

\vspace{0.5em}
\noindent\textbf{\textit{Lemma 7}}: Given a set
of logs collected from all servers, the auditor detects
all incomplete logs -- logs with missing tail entries.

\textit{\textbf{Proof}}:
A subset of servers cannot successfully 
modify arbitrary blocks in the log (proof in Lemma 6)
but they can omit the tail of the log. During an audit, 
the auditor gathers the logs from all the
servers. At least
one correct server exists with the complete log -- which
can easily be verified for correctness
by validating the collective signature and
hash pointer in each block. The auditor uses
this complete and verified log to detect that one or
more servers store an incomplete log.\hfill$\Box$
\subsection{Correctness of \system}
\label{sub:correct}

\noindent\textit{\textbf{Definition 1}: Verifiable ACID properties}

In transaction processing, ACID refers to the four 
key components of a transaction: \\
i) \underline{A}tomicity: A transaction is an atomic unit in that
either all operations are executed or none.\\
ii) \underline{C}onsistency: Data is in a consistent
state before and after a transaction executes.\\
iii) \underline{I}solaiton: When transactions are executed
concurrently, isolation ensures that the transactions
seem to have executed sequentially.\\
iv) \underline{D}urability: If a transaction commits,
its updates are persistent even in the presence of
failures.

% Traditional trust-based DBMS provide these guarantees
% as is but
We define \textit{v-ACID} as the ACID properties 
that can be verified. \textit{v-ACID} indicates that
a database system provides verifiable evidence
that the ACID guarantees are upheld. 
% an external entity can verify whether a database system
% provides ACID guarantees. 
This definition is 
useful when individual database servers are
untrusted and may violate ACID -- in which case the
system must allow verifying and detecting the
violations.

\vspace{0.5em}
\noindent\textit{\textbf{Theorem 1}: \system provides Verifiable
ACID guarantees.}\\
\textit{\textbf{Proof}}: \system guarantees that an external 
auditor can verify if the database servers provide
ACID guarantees or not. 

The first step in the verification
is for the auditor to obtain a \textit{correct} and
\textit{complete} log. Given the assumption that
at least one server is correct at a given time,
Lemmas 6 and 7 prove that during an audit, the
auditor always identifies the correct and complete log.

Lemma 5 proves that \textit{\underline{A}tomicity}
violation is 
verifiable; Lemma 2 proves that the
auditor verifies if the effect of a transaction
resulted in an inconsistent database when a server
buffers inconsistent writes, i.e., verifiable
\textit{\underline{C}onsistency}; Lemma 3 proves that 
the \textit{\underline{I}solation}
guarantee which ensures serializable transaction 
execution is verifiable; and finally,
Lemmas 1 and 2 verify
if the effects of committed transactions are
\textit{\underline{D}urable}. Hence, an auditor
verifies whether the servers in \system uphold ACID 
properties.

Note that multiple ACID violations can exist in
the transaction execution. Since the log is sequential,
the auditor identifies the first occurrence of any of 
these violations and the blocks after that need not be 
audited since everything following that violation
can be incorrect and hence irrelevant to a
correct execution.\hfill$\Box$

\subsection{Scaling \protocol protocol}
\label{sub:scaling}

The \protocol protocol discussed in Section~\ref{sub:tcl}
makes simplifying assumptions that each block contains a single
transaction and a globally designated coordinator terminates
all transactions which requires participation fromm
all servers.
This makes \protocol expensive as any server
not involved in a transaction must also participate in its
termination. In this section we provide an intuitive
overview of how to scale \protocol. 

To scale \protocol, two aspects can be enhanced: (i) Allow
multiple transactions to commit simultaneously by
storing multiple transactions in a block, and (ii)
Reduce the number of
servers participating in transaction termination to
only the servers involved in that transaction.

Extending each block to contain multiple transactions
is straight-forward. The coordinator collects and
inserts a set of \textit{non-conflicting}
client generated transactions and orders 
them within a single
block at the start of \protocol. Once the protocol 
begins, the coordinator or any other server cannot 
re-order the transactions within the block (the argument
is similar to Lemma 4). This technique allows each 
execution of \protocol to commit multiple transactions. 
In our evaluations in Section~\ref{sec:eval},
we store multiple transactions in each block.

To reduce the number of servers participating in
transaction termination, servers are divided
into small dynamic \textit{groups}. The servers
accessed by a transaction forms one group, in which
one server acts as the coordinator to terminate
that transaction
(instead of one globally designated coordinator).
Each group executes \protocol 
internally and upon a successful execution, the coordinators
of each group publish the block to all other
groups. The problem with such a solution is in deciding
the order of blocks \textit{across} groups
such that all the servers maintain a consistently
ordered transaction log.

\begin{figure}[h!]
\centering
\includegraphics[scale=0.4]{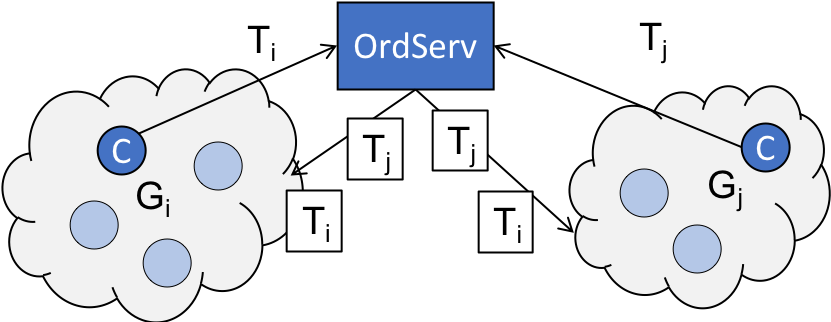}
\caption{Scaling \protocol.}
\label{fig:scaling_2}
\vspace{-1em}
\end{figure}

There are multiple ways to solve the ordering problem.
Figure~\ref{fig:scaling_2} depicts a scalable solution
that abstracts the ordering of blocks as a service (OrdServ). 
The figure shows two groups of servers $G_i$ and $G_j$, 
each accessed by transactions $T_i$ and $T_j$
respectively. The OrdServ component is responsible for
atomically broadcasting a single stream of blocks, each
generated by \protocol executed in different 
groups of servers. OrdServ can use
a byzantine consensus protocol such as PBFT
\cite{castro1999practical} among the coordinators to
consistently order blocks; or it can be an 
off-the-shelf application such as Apache Kafka,
used to provide ordering service in a recent work, 
Veritas~\cite{allen2019veritas}.
OrdServ is also responsible 
for chaining the blocks i.e., the 
coordinators of the groups do not fill in the hash of
previous block, rather it is filled by the OrdServ.
There are two possible scenarios regarding the groups:

%\begin{itemize}
$\bullet$ $G_i \cap G_j = \varnothing$: If any two groups 
    of servers have no overlapping server, there is no 
    dependency between the two blocks of transactions
    $T_i$ and $T_j$, and OrdServ can order them in any way and
    broadcast a consistent order.
    
$\bullet$ $G_i \cap G_j \neq \varnothing$: If any two groups
    have a non-empty intersection,
    then transactions $T_i$ and $T_j$ may have a dependency order (e.g., 
    $T_j$ wrote a data item after $T_i$ read it); the OrdServ
    should ensure that the
    transaction log reflects this dependency between the published blocks.
%\end{itemize}

Although there is flexibility in choosing OrdServ, it is 
important to choose a solution that maintains local 
transaction order (within a group) across the globally 
replicated log. Solutions such a ParBlock
\cite{amiri2019parblockchain} track the transaction 
dependency order and maintains that order while 
publishing blocks. We plan to integrate 
ParBlock with
\protocol as future work.

\section{Failure Examples}
\label{sec:failures}

In this section we discuss various malicious failures 
and safety violation scenarios and explain how the
failures are detected. The failure model of \system
permits a server to misbehave but captures enough 
details in the transaction log for an auditor
to detect the malicious failures as well as the failing
servers.

\vspace{0.5em}
\noindent\textbf{\textit{Scenario 1: Incorrect Reads}}

A malicious server can respond with incorrect
values for the data items read in the read
requests. We use Lemma 1 to detect this.

\begin{figure}[h!]
\centering
\includegraphics[scale=0.3]{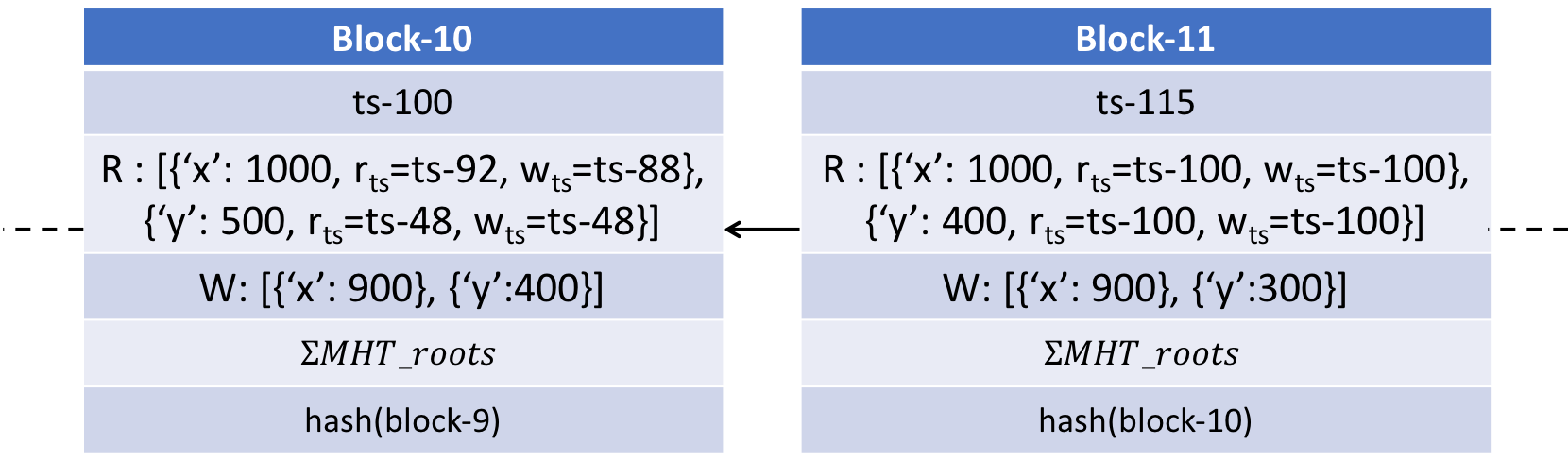}
\caption{Isolation guarantee violation example.}
\label{fig:violation_ex}
\vspace{-0.5em}
\end{figure}

Figure~\ref{fig:violation_ex} gives an example of 
incorrect reads.
Assume that the severs store bank details and
there are two transactions $T_1$ and $T_2$ deducting 
\$100 from two accounts, $x$ and $y$.
Block-10 contains $T_1$ and
Block-11 contains $T_2$. $T_1$
reads two data items: one with id $x$, value 1000, $r_{ts}$ = 
\textit{ts-92}, and $w_{ts}$ = \textit{ts-88}, and the second with id $y$, value 500, $r_{ts}$ = \textit{ts-48}, and
$w_{ts}$ = \textit{ts-48}. $T_1$
updates $x$ to \$900 and $y$ to \$400, and upon commitment,
it also updates their $r_{ts}$ and $w_{ts}$ to
\textit{ts-100}. Any transaction executing after this
must reflect the latest data. But $T_2$, 
committing at timestamp \textit{ts-115}, 
has incorrect value of \$1000 for $x$ (but up-to-date
timestamps).
This indicates that the server
storing data items $x$ is misbehaving by sending
incorrect read values.

\vspace{0.5em}
\noindent\textbf{\textit{Scenario 2: Incorrect 
Block Creation}}

While executing \protocol to terminate a transaction 
$T_i$, a malicious coordinator can
add an incorrect Merkle Hash Tree (MHT) root of a benign 
server $S_b$ in the block; this can cause audit
failure of $S_b$ (as 
Lemma 2 uses MHT roots to detect datastore corruption).
But such an attempt will be detected by
the benign server, as proved in Lemma 5. 

In the \textit{vote} phase of \protocol, explained
in Section~\ref{sub:tcl}, server
$S_b$ sends the MHT root
corresponding to transaction $T_i$
to the coordinator. If the coordinator stores an incorrect
MHT root or a correct root but corresponding to an
older transaction $T_{i-1}$, $S_b$ can detect this in the
$schResponse$ phase of \protocol.
and not cooperate to produce a valid
co-sign.

\vspace{0.5em}
\noindent\textbf{\textit{Scenario 3: Data corruption}}

A server may corrupt the data stored in the
datastore, essentially not reflecting the expected 
changes requested by the clients. We assume a multi-versioned
datastore in this example and use
Verification Objects \vo and MHT roots to detect datastore
corruption, as proved in Lemma 2.
\begin{figure}[t!]
\centering
\includegraphics[scale=0.38]{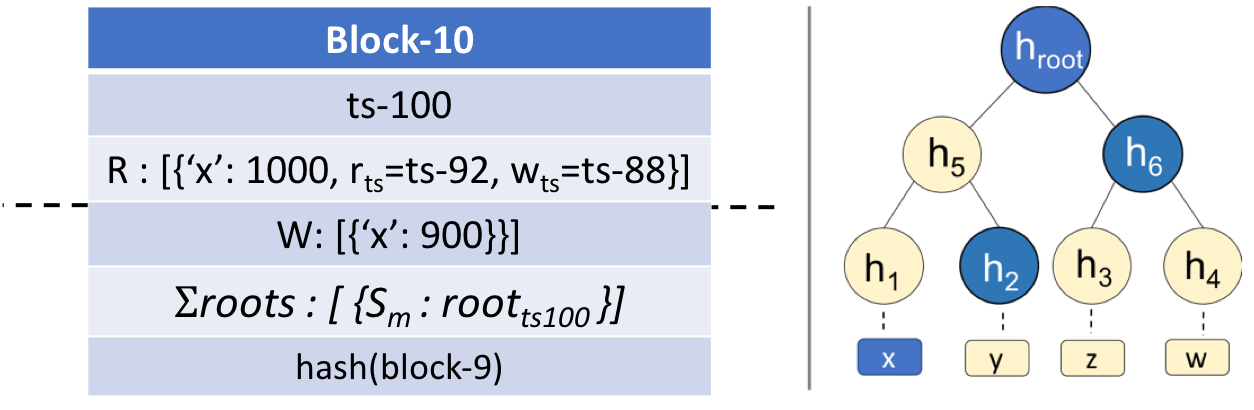}
\caption{Data corruption example}
\label{fig:ds_layer}
\vspace{-1em}
\end{figure}
Consider a transaction $T_i$ committed
at timestamp \textit{ts-100} and updated a data item $x$
stored in $S_m$. 
Figure~\ref{fig:ds_layer} indicates the
data stored in server $S_m$ that is being audited at
version \textit{ts-100}. The auditor
fetches the corresponding block (block 10) from the 
log and extracts $x$'s value written by $T_i$ and the MHT root 
corresponding to $S_m$. This MHT root should reflect $x$'s
updated value.

Assume $S_m$ was malicious and did not update
$x$ to 900. In the next step of verification, auditor
asks $S_m$ for the \vo of data item $x$ at timestamp \textit{ts-100}.
$S_m$ responds with \{$h_2, h_6, h_{root}$\} 
(hash values of the sibling nodes of data $x$ in the path
from leaf
to root). Auditor hashes $x$'s value stored in the block ($H(900)$) and uses
$h_2$ sent in \vo to compute $h_5^\prime$ and further, hash 
$h_5^\prime$ and $h_6$ (from \vo) to compute the expected root,
$h_{root}^\prime$.
This computed root should match the root 
the root stored in the block i.e.,
$h_{root}^\prime = root_{S_m-ts100}$. But since $S_m$ did
not update the value of $x$ to 900, the root computed
by the auditor will not
not match the root stored in the block (assuming collision-free
hash functions). Thus data corruption at $S_m$, precisely
at version \textit{ts-100} is detected.
\section{Evaluation}
\label{sec:eval}

In this section, we discuss the experimental evaluation
of \protocol. Our goal is to 
measure the overhead incurred 
in executing an atomic commit protocol on untrusted 
infrastructure. The focus of \system and \protocol
is \textit{fault detection} in a non-replicated system, 
hence solutions based on replication that typically use
PBFT~\cite{castro1999practical} are orthogonal to 
\protocol.

In evaluating \protocol, we measure the 
performance using two 
aspects: \textit{commit latency} - time taken to terminate a
transaction once the client sends \textsf{end transaction}
request, and \textit{throughput} - the number of transactions
committed per second; \protocol was implemented in Python. 
We deployed multiple
database servers on a single
Amazon AWS datacenter (US-West-2 region) where each server was
an
\textsf{EC2 m5.xlarge} vm consisting of 4 vCPUs, 16 GiB RAM
and upto 10 Gbps network bandwidth. Unless otherwise specified 
in the experiment, each database server stores a single shard
(or partition) of data consisting of 10000 data items.

To evaluate the protocol, we used Transactional-YCSB-like
benchmark~\cite{cooper2010benchmarking} consisting of
transactions with read-write
operations. Each transaction consisted of 5 operations on
different data items thus generating a multi-record workload.
The data items were picked at random from a pool of all the
data partitions combined, resulting in distributed transactions.
Although we presented \protocol and \system with the simplifying
assumption of one transaction per block, in the experiments,
we typically stored 100 non-conflicting transactions
in each block.
Every experimental run consisted of 1000 client requests and
each data point plotted in this section is an average of
3 runs. 

\subsection{\protocol vs. 2PC}
As a first step, we compare
the trust-free protocol \protocol
with its trusted counterpart 
Two Phase Commit~\cite{gray1978notes}.
\protocol is essentially 2PC combined
with the cryptographic
primitives (Co-Signing and Merkle Hash Trees) which results
in an additional phase due to the trust-free nature. Thus,
comparing \protocol with 2PC highlights the overhead 
incurred
by \protocol to operate in an untrusted setting. 
Both 2PC and \protocol are implemented such that 
transactions are terminated and blocks are produced 
\textit{sequentially}  so that the log does not
have forks.

\begin{figure}[ht!]
\begin{center}
\resizebox{0.3\textwidth}{0.2\textwidth}
{% This file was created by matplotlib2tikz v0.6.18.
\begin{tikzpicture}

\definecolor{color1}{rgb}{0.91, 0.45, 0.32}
\definecolor{color0}{rgb}{0.415686274509804,0.352941176470588,0.803921568627451}

\begin{axis}[
axis y line=right,
%legend cell align={left},
legend entries={{TFC Throughput},{2PC Throughput}},
legend style={at={(1,1)}},
legend style={draw=white!80.0!black},
legend style={font=\normalsize},
tick align=outside,
tick pos=right,
x grid style={lightgray!92.02614379084967!black},
xlabel style={font=\LARGE},
xmin=2.68, xmax=5.52,
xtick={0.1,1.1,2.1},
xticklabels={3,5,7},
xticklabel style = {font=\large},
y grid style={lightgray!92.02614379084967!black},
ylabel style={font=\LARGE},
ylabel={Throughput (txns/sec)},
yticklabel pos=right,
yticklabel style = {font=\Large},
ymin=0, ymax=700
]
\addlegendimage{ybar,ybar legend,fill=color1,draw opacity=1,fill opacity=0.6,
postaction={
        pattern=dots
    }};
\draw[fill=color0,draw opacity=1,fill opacity=0.6, postaction={
        pattern=crosshatch
    }] (axis cs:2.9,0) rectangle (axis cs:3.0,513.11);
\draw[fill=color0,draw opacity=1,fill opacity=0.6, postaction={
        pattern=crosshatch
    }] (axis cs:4.0,0) rectangle (axis cs:4.1,369.14);
\draw[fill=color0,draw opacity=1,fill opacity=0.6, postaction={
        pattern=crosshatch
    }] (axis cs:5.1,0) rectangle (axis cs:5.2,286.52);
\addlegendimage{ybar,ybar legend,fill=color0,draw opacity=1,fill opacity=0.6,
postaction={
        pattern=crosshatch
    }};
\draw[fill=color1,draw opacity=1,fill opacity=0.6, postaction={
        pattern=dots
    }] (axis cs:3.0,0) rectangle (axis cs:3.1,263.44);
\draw[fill=color1,draw opacity=1,fill opacity=0.6, postaction={
        pattern=dots
    }] (axis cs:4.1,0) rectangle (axis cs:4.2,206.99);
\draw[fill=color1,draw opacity=1,fill opacity=0.6, postaction={
        pattern=dots
    }] (axis cs:5.2,0) rectangle (axis cs:5.3,161.1);
\path [draw=lightgray!40.0!black, semithick] (axis cs:3,508.365472995363)
--(axis cs:3,517.854527004637);

\path [draw=lightgray!40.0!black, semithick] (axis cs:5,366.8672690177)
--(axis cs:5,371.4127309823);

\path [draw=lightgray!40.0!black, semithick] (axis cs:7,284.42548106514)
--(axis cs:7,288.61451893486);

\path [draw=lightgray!40.0!black, semithick] (axis cs:3.2,259.44)
--(axis cs:3.2,267.44);

\path [draw=lightgray!40.0!black, semithick] (axis cs:5.2,204.99)
--(axis cs:5.2,208.99);

\path [draw=lightgray!40.0!black, semithick] (axis cs:7.2,160.1)
--(axis cs:7.2,162.1);

\end{axis}

\begin{axis}[
axis y line=left,
legend style={at={(1,1)}},
legend entries={{TFC Latency},{2PC Latency}},
legend style={at={(0.03,0.97)}, anchor=north west, draw=white!80.0!black},
legend style={font=\normalsize},
tick align=outside,
tick pos=left,
x grid style={lightgray!92.02614379084967!black},
xlabel style={font=\LARGE},
xlabel={Number of servers},
xticklabel style = {font=\Large},
% xmin=2.68, xmax=5.52,
% xtick pos=left,
% xtick={0.1,1.1,2.1},
% xticklabels={3,5,7},
y grid style={lightgray!92.02614379084967!black},
ylabel style={font=\LARGE},
yticklabel pos=left,
yticklabel style = {font=\Large},
ylabel={Latency (ms)},
ymin=0.727, ymax=8.413
]
\addlegendimage{mark=*, color1, mark options={solid,draw=black}}
\addlegendimage{mark=star, color0, dashed, mark options={solid,draw=black}}
\addplot [thick, color0, dashed, mark=star, mark size=2, mark options={draw=black}]
table [row sep=\\]{%
3	1.94 \\
5	2.71 \\
7	3.49 \\
};
\addplot [thick, color1, mark=*, mark size=1.5, mark options={solid,draw=black}]
table [row sep=\\]{%
3	3.79 \\
5	4.83 \\
7	6.2 \\
};
\end{axis}

\end{tikzpicture}}
\caption{2PC vs. \protocol (TFC).}
\label{fig:2pc_vs_tfc}
\end{center}
\vspace{-1em}
\end{figure}

Figure~\ref{fig:2pc_vs_tfc} contrasts the performance of
2PC vs. \protocol. We increase the number of servers and 
measure commit latency and throughput. In this experiment, 
each block stores
\textit{a single} transaction so that we can measure
the overhead induced by \protocol \textit{per transaction}.
Given that
each block contains a single transaction and that blocks are
generated sequentially, the servers are essentially committing
one transaction after another.

As indicated in the figure, the average latency to commit a
single transaction in an untrusted setting is approximately
\textsf{1.8x}
more than a trusted environment. The throughput for 2PC
is approximately \textsf{2.1x} higher than \protocol. 
\protocol performs additional computations
compared with 2PC: Merkle Hash Tree (MHT) updates to
compute new roots after each transaction, collective
signature on each block, and an additional phase. 
In spite of the additional computing and achieving 
trust-free atomic commitment, \protocol is only
\textsf{1.8x} slower than 2PC. Having shown the 
overhead of \protocol as compared to 2PC,
the following experiments measure the performance of
\protocol by varying different parameters.

\subsection{Number of transactions per block}
\label{sub:inc_txns}
\begin{figure}[ht!]
\begin{center}
\resizebox{0.3\textwidth}{0.2\textwidth}
{% This file was created by matplotlib2tikz v0.6.18.
\begin{tikzpicture}

\definecolor{color0}{rgb}{0.83921568627451,0.152941176470588,0.156862745098039}

\begin{axis}[
axis y line=right,
legend cell align={left},
legend entries={{Throughput (tps)}, {Latency (ms)}},
legend style={at={(0.03,0.97)}, anchor=north west, draw=white!80.0!black},
legend style={font=\normalsize},
tick align=outside,
x grid style={lightgray!92.02614379084967!black},
xlabel style={font=\LARGE},
xmin=-5.55, xmax=127.55,
xtick pos=left,
xticklabels={1, 2, 20, 40, 60, 80, 100, 120},
y grid style={lightgray!92.02614379084967!black},
ylabel style={font=\LARGE},
ylabel={Throughput (txns/sec)},
ymin=200, ymax=1653.24019899029,
yticklabel style = {font=\Large},
xticklabel style = {font=\Large},
ytick pos=right
]
\addlegendimage{ybar,ybar legend,fill=blue,draw opacity=1,fill opacity=0.5};
\addlegendimage{mark=*, color0, mark options={solid,draw=black}}
\draw[fill=blue,draw opacity=1,fill opacity=0.5] (axis cs:0.5,0) rectangle (axis cs:3.5,360.73050464382);
\draw[fill=blue,draw opacity=1,fill opacity=0.5] (axis cs:4.5,0) rectangle (axis cs:7.5,749.400445224328);
\draw[fill=blue,draw opacity=1,fill opacity=0.5] (axis cs:8.5,0) rectangle (axis cs:11.5,947.439098267654);
\draw[fill=blue,draw opacity=1,fill opacity=0.5] (axis cs:12.5,0) rectangle (axis cs:15.5,1063.15569373062);
% \draw[fill=blue,draw opacity=1,fill opacity=0.5] (axis cs:16.5,0) rectangle (axis cs:19.5,1146.0682633329);
\draw[fill=blue,draw opacity=1,fill opacity=0.5] (axis cs:18.5,0) rectangle (axis cs:21.5,1171.70121023779);
\draw[fill=blue,draw opacity=1,fill opacity=0.5] (axis cs:38.5,0) rectangle (axis cs:41.5,1290.42038091471);
\draw[fill=blue,draw opacity=1,fill opacity=0.5] (axis cs:58.5,0) rectangle (axis cs:61.5,1321.3118149523);
\draw[fill=blue,draw opacity=1,fill opacity=0.5] (axis cs:78.5,0) rectangle (axis cs:81.5,1336.41923713361);
\draw[fill=blue,draw opacity=1,fill opacity=0.5] (axis cs:98.5,0) rectangle (axis cs:101.5,1337.97459462782);
\draw[fill=blue,draw opacity=1,fill opacity=0.5] (axis cs:118.5,0) rectangle (axis cs:121.5,1338.58201724084);
\end{axis}

\begin{axis}[
legend style={font=\normalsize},
tick align=outside,
tick pos=left,
x grid style={lightgray!92.02614379084967!black},
xlabel style={font=\LARGE},
xlabel={Number of txns per block},
xmin=-5.55, xmax=127.55,
y grid style={lightgray!92.02614379084967!black},
xticklabels={1, 2, 20, 40, 60, 80, 100, 120},
ylabel style={font=\LARGE},
yticklabel style = {font=\Large},
xticklabel style = {font=\Large},
ylabel={Latency (ms)},
ymin=0.647074031005995, ymax=3.49335304385278
]
\addplot [thick, color0, mark=*, mark size=1.5, mark options={solid,draw=black}, forget plot]
table [row sep=\\]{%
2	2.77215854326883 \\
6	1.33440374003513 \\
10	1.05548733075455 \\
14	0.94059627964381 \\
18	0.872550412460607 \\
20	0.853465983361903 \\
40	0.774946044361773 \\
60	0.75683160879278 \\
80	0.74826853158994 \\
100	0.765717021000927 \\
120	0.749860301043813 \\
};
\end{axis}

\end{tikzpicture}}
\caption{Varying number of transaction per block}
\label{fig:txns_per_block}
\end{center}
\vspace{-0.8em}
\end{figure}
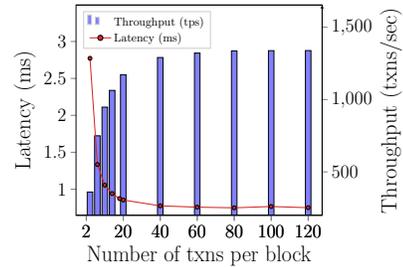

In this experiment, we fix the number of servers to 5 
and increase the load on the system by increasing 
the number of transactions stored within each block. 
Each database server consisted of 10000 data items.
Figure~\ref{fig:txns_per_block}
indicates the average latency to commit a single transaction
and the throughput while increasing number of transactions
stored within each block from 2 to 120. The latency to commit a
single transaction reduces by \textsf{2.6x} and the throughput
increases by \textsf{2.5x} when 80 or more transactions are
batched in a single block. This experiment highlights that
even though the blocks are produced sequentially, the 
performance of \protocol can be significantly enhanced by 
processing multiple transactions in one block.

\subsection{Number of shards}

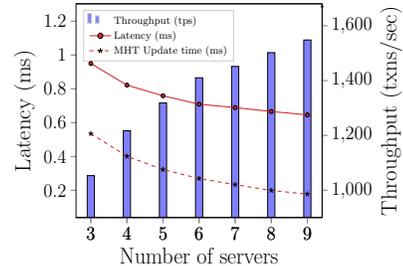
\begin{figure}[ht!]
\begin{center}
\resizebox{0.3\textwidth}{0.2\textwidth}
{% This file was created by matplotlib2tikz v0.6.18.
\begin{tikzpicture}

\definecolor{color0}{rgb}{0.83921568627451,0.152941176470588,0.156862745098039}
\definecolor{color1}{rgb}{1,0.647058823529412,0}

\begin{axis}[
axis y line=right,
legend cell align={left},
legend entries={{Throughput (tps)}, {Latency (ms)}, {MHT Update time (ms)}},
legend style={at={(0.03,0.97)}, anchor=north west, draw=white!80.0!black},
legend style={font=\small},
tick align=outside,
x grid style={lightgray!92.02614379084967!black},
xlabel style={font=\LARGE},
xmin=2.59, xmax=9.41,
xtick pos=left,
y grid style={lightgray!92.02614379084967!black},
ylabel style={font=\LARGE},
ylabel={Throughput (txns/sec)},
ymin=900, ymax=1680.23419068817,
yticklabel style = {font=\Large},
xticklabel style = {font=\Large},
ytick pos=right
]
\addlegendimage{ybar,ybar legend,fill=blue,draw opacity=1,fill opacity=0.5};
\addlegendimage{mark=*, color0, mark options={solid,draw=black}}
\addlegendimage{mark=star, dashed, color0, mark options={solid,draw=black}}
\draw[fill=blue,draw opacity=1,fill opacity=0.5] (axis cs:2.9,0) rectangle (axis cs:3.1,1052.98354312836);
\draw[fill=blue,draw opacity=1,fill opacity=0.5] (axis cs:3.9,0) rectangle (axis cs:4.1,1216.80544965192);
\draw[fill=blue,draw opacity=1,fill opacity=0.5] (axis cs:4.9,0) rectangle (axis cs:5.1,1318.28440230617);
\draw[fill=blue,draw opacity=1,fill opacity=0.5] (axis cs:5.9,0) rectangle (axis cs:6.1,1410.23894734538);
\draw[fill=blue,draw opacity=1,fill opacity=0.5] (axis cs:6.9,0) rectangle (axis cs:7.1,1451.65161017921);
\draw[fill=blue,draw opacity=1,fill opacity=0.5] (axis cs:7.9,0) rectangle (axis cs:8.1,1502.00790157433);
\draw[fill=blue,draw opacity=1,fill opacity=0.5] (axis cs:8.9,0) rectangle (axis cs:9.1,1548.15891441199);
\end{axis}

\begin{axis}[
legend style={font=\small},
tick align=outside,
tick pos=left,
x grid style={lightgray!92.02614379084967!black},
xlabel style={font=\LARGE},
xlabel={Number of servers},
xmin=2.59, xmax=9.41,
y grid style={lightgray!92.02614379084967!black},
ylabel style={font=\LARGE},
ylabel={Latency (ms)},
yticklabel style = {font=\Large},
xticklabel style = {font=\Large},
ymin=0.038640412092208, ymax=1.302724672675136,
]
\addplot [thick, color0, mark=*, mark size=1.5, mark options={solid,draw=black}, forget plot]
table [row sep=\\]{%
3	0.949684083461757 \\
4	0.82182631174723 \\
5	0.758566912015277 \\
6	0.70910236835479 \\
7	0.688872299194173 \\
8	0.665777022043857 \\
9	0.645930776596043 \\
};
\addplot [thick, color0, dashed, mark=star, mark size=2, mark options={solid,draw=black}, forget plot]
table [row sep=\\]{%
3	0.534753175576493 \\
4	0.4012767362593 \\
5	0.322399755318959 \\
6	0.269489013353983 \\
7	0.233450872103373 \\
8	0.198865916728973 \\
9	0.177261539300281 \\
};
\end{axis}

\end{tikzpicture}}
\caption{Varying number of servers.}
\label{fig:num_shards}
\end{center}
\vspace{-0.8em}
\end{figure}

In this experiment, we measure the scalability
of \protocol by increasing the number of database servers (each
storing a shard of 10000 data items) from 3 to 9, while keeping
the number of transaction per block constant (100 per block).
Figure
\ref{fig:num_shards} depicts the experimental results. The throughput of \protocol increases
by 47\% and the commit latency reduces by 33\% when the
number of servers are increased from 3 to 9. 
Figure~\ref{fig:num_shards} also shows the most expensive operation in
committing transactions i.e., Merkle Hash Tree (MHT) updates.
Recall from Section~\ref{sub:tcl} that in
\protocol, termination of each transaction requires computing
the updated MHT root.
Given that each block has 100 transactions, which in turn
consists of 5 operations each, there are 500 operations in
each block. With only 3 servers, all the operations access the 
three shards whereas with 9 servers, the 500 operations are
spread across nine shards. Thus, the load per server reduces
when there are more servers, resulting in the
reduction of MHT update latencies. This experiment highlights
that \protocol is scalable and performs well with
increasing number of database servers.

\subsection{Number of data items}

\begin{figure}[ht!]
\begin{center}
\resizebox{0.3\textwidth}{0.2\textwidth}
{% This file was created by matplotlib2tikz v0.6.18.
\begin{tikzpicture}

% \definecolor{blue}{rgb}{0.91, 0.45, 0.32}
% \definecolor{color0}{rgb}{0.415686274509804,0.352941176470588,0.803921568627451}
\definecolor{color0}{rgb}{0.83921568627451,0.152941176470588,0.156862745098039}

\begin{axis}[
axis y line=right,
legend cell align={left},
legend entries={{Throughput (tps)}, {Latency}},
legend style={draw=white!80.0!black},
legend style={font=\normalsize},
tick align=outside,
x grid style={lightgray!92.02614379084967!black},
xlabel style={font=\LARGE},
xmin=549.945, xmax=10450.055,
xtick pos=left,
xtick={1000,2000,3000,4000,5000,6000,7000,8000,9000,10000},
xticklabels={1k, 2k, 3k, 4k, 5k, 6k, 7k, 8k, 9k, 10k},
y grid style={lightgray!92.02614379084967!black},
ylabel={Throughput (txns/sec)},
ylabel style={font=\LARGE},
ymin=1100, ymax=1694.99610700759,
yticklabel style = {font=\Large},
xticklabel style = {font=\Large},
ytick pos=right
]
% \draw[fill=color0,draw opacity=1,fill opacity=0.5, postaction={
%         pattern=crosshatch
%     }] (axis cs:2.9,0) rectangle (axis cs:3.1,513.11);
    
\addlegendimage{ybar,ybar legend,fill=blue,draw opacity=1,fill opacity=0.5};
\addlegendimage{mark=*, color0, mark options={solid,draw=black}}
\draw[fill=blue,draw opacity=1,fill opacity=0.5] (axis cs:850,0) rectangle (axis cs:1150.1,1566.66295905485);
\draw[fill=blue,draw opacity=1,fill opacity=0.5] (axis cs:1850,0) rectangle (axis cs:2150.1,1507.57903735828);
\draw[fill=blue,draw opacity=1,fill opacity=0.5] (axis cs:2850,0) rectangle (axis cs:3150.1,1477.79783130569);
\draw[fill=blue,draw opacity=1,fill opacity=0.5] (axis cs:3850,0) rectangle (axis cs:4150.1,1451.36666560715);
\draw[fill=blue,draw opacity=1,fill opacity=0.5] (axis cs:4850,0) rectangle (axis cs:5150.1,1401.55417569831);
\draw[fill=blue,draw opacity=1,fill opacity=0.5] (axis cs:5850,0) rectangle (axis cs:6150.1,1376.87947346483);
\draw[fill=blue,draw opacity=1,fill opacity=0.5] (axis cs:6850,0) rectangle (axis cs:7150.1,1375.44612277613);
\draw[fill=blue,draw opacity=1,fill opacity=0.5] (axis cs:7850,0) rectangle (axis cs:8150.1,1370.66859107638);
\draw[fill=blue,draw opacity=1,fill opacity=0.5] (axis cs:8850,0) rectangle (axis cs:9150.1,1358.72878486727);
\draw[fill=blue,draw opacity=1,fill opacity=0.5] (axis cs:9850,0) rectangle (axis cs:10150.1,1348.61603495837);
\end{axis}

\begin{axis}[
legend style={font=\Large},
tick align=outside,
tick pos=left,
x grid style={lightgray!92.02614379084967!black},
xlabel style={font=\LARGE},
xlabel={Number of data items per shard},
xmin=549.945, xmax=10450.055,
xtick pos=left,
xtick={1000,2000,3000,4000,5000,6000,7000,8000,9000,10000},
xticklabels={1k, 2k, 3k, 4k, 5k, 6k, 7k, 8k, 9k, 10k},
y grid style={lightgray!92.02614379084967!black},
ylabel style={font=\LARGE},
ylabel={Latency (ms)},
ymin=0.58, ymax=0.78,
ytick={0.6,0.65,0.7,0.75},
yticklabel style = {font=\Large},
xticklabel style = {font=\Large},
yticklabels={0.60,0.65,0.70,0.75}
]
\addplot [thick, color0, mark=*, mark size=1.5, mark options={solid,draw=black}, forget plot]
table [row sep=\\]{%
1000    0.638311179478957 \\
2000    0.6633186173439 \\
3000    0.67229887644449 \\
4000    0.69539617379506 \\
5000    0.71625809033708 \\
6000    0.72628045002619 \\
7000    0.727052810192103 \\
8000    0.73271747986475 \\
9000    0.736055735747017 \\
10000   0.738374629020687 \\
};
\end{axis}

\end{tikzpicture}}
\caption{Varying number of data items per shard}
\label{fig:items_per_shard}
\end{center}
\vspace{-0.8em}
\end{figure}
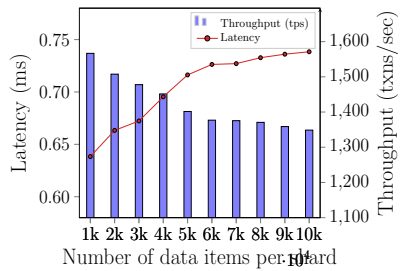

In the final set of experiments, we measure the performance of
\protocol
by varying the number of data items stored in each database
server, while keeping a constant of 100 transactions per block
and using 5 database servers. The number of
items stored in each server increased from 1000 to 10000
to measure the commit latency and throughput of \protocol,
as shown in Figure~\ref{fig:items_per_shard}.
The commit latency increases by 15\% and the throughput reduces
by 14\% with the increase in number of data items per shard.
The performance fluctuation is due to the Merkle Hash Tree
updates that varies with the number of data items. Updating a 
single leaf node in a binary
hash tree with 1000 leaf nodes (data items) updates 10 
nodes (from leaf to the root) and a tree with 10000 leaf nodes
updates roughly 14 nodes. Thus, the performance of \protocol
decreases with increasing number of data items stored within
each server.
\section{Related Work}
\label{sec:related_work}
The literature on databases that tolerate
malicious failures is extensive
\cite{garcia1986applications, gashi2004designing, vandiver2007tolerating, garcia2011efficient, luiz2011byzantine, pedone2012byzantine}.
All of these solutions differ from \textit{\system} as they:
assume a singe non-partitioned database, rely on \textit{replicating} the database
to tolerate byzantine failures, and some
also require a trusted component for correctness. Garcia-Molina et al.\cite{garcia1986applications} were the
earliest to propose a set of
database schemes that tolerate malicious faults. The
work presents the theoretical foundations on replicating
the database on enough servers to handle malicious faults 
but lacks a practical implementation. Gashi et al.~\cite{gashi2004designing}
discuss fault-tolerance other than just
crash failures and provide a report composed of database failures
caused by software bugs. HRDB by Vandiver et al.
\cite{vandiver2007tolerating} propose a replication scheme to 
handle byzantine faults wherein a trusted coordinator delegates
transactions to the replicas. The coordinator also orders the
transactions and decides when to safely commit a transaction.
Byzantium by Garcia et al.~\cite{garcia2011efficient}
provides an efficient replicated middleware between the client 
and the database to tolerate byzantine faults. It differs from
previous solutions by allowing concurrent transactions and by
not requiring a trusted component to coordinate the replicas. 

The advent of blockchains brought with it a set
of technologies that manage data in untrusted 
environments. In both the open perimissionless and closed permissioned 
blockchains, due to lack of trust,
the underlying protocols must be designed
to tolerate any type of malicious behavior.  But these
protocols and their applications are mostly limited to
crypto-currencies and cannot be easily extended for
large scale distributed data management. 
Although permissionless blockchain solutions
such as Elastico~\cite{luu2016secure} 
Omniledger~\cite{kokoris2018omniledger}, and
RapidChain~\cite{zamani2018rapidchain} discuss 
sharding, it is with respect to transactions, i.e.,
different servers execute different transactions to
enhance performance but all of them maintain copies of same
data, essentially acting as replicas of
a single database. These solutions differ from 
\system as they focus
of replicated data rather than distributed data.

In the space of transaction commitment,
proposals such as ~\cite{mohan1983method,zhao2007byzantine, al2017chainspace,zhang2011trustworthy} tolerate malicious 
faults. Mohan et al.~\cite{mohan1983method}
integrated 2PC with byzantine fault-tolerance to make 2PC
non-blocking and to prevent the coordinator from
sending conflicting decisions. Zhao et al.\cite{zhao2007byzantine} propose a commit protocol that
tolerates byzantine faults at the coordinator by replicating
it on enough servers to run a byzantine agreement
protocol to agree on the transaction decision. Chainspace
\cite{al2017chainspace} proposes a commit protocol in a 
blockchain setting wherein each shard is replicated on
multiple servers to allow executing byzantine 
agreement per shard to agree on the transaction decision.
All these solutions require replication and execute
byzantine agreement on the replicas,
and hence differ from \protocol. \protocol uses Collective
Signing (CoSi)~\cite{syta2016keeping}, a cryptographic multisignature
scheme to tolerate malicious failures during commitment. CoSi has been adapted
to make consensus more efficient in blockchains,
e.g., ByzCoin~\cite{kogias2016enhancing}. To
our knowledge, \protocol is the first to merge
CoSi with atomic commitment.

 \system uses a tamper-proof log to audit the system and detect
any failures across database servers; this technique
has been studied for decades in distributed systems
\cite{yumerefendi2004trust, yumerefendi2005role, yumerefendi2007strong, haeberlen2007peerreview}.
In~\cite{yumerefendi2004trust} and
\cite{yumerefendi2005role}, Yumerefendi et al. highlight
the use of accountability -- a mechanism to detect and
expose misbehaving servers-- as a general distributed
systems design. They implement CATS~\cite{yumerefendi2007strong} an accountable network storage
system that uses secure
message logs to detect and expose misbehaving nodes. PeerReview
\cite{haeberlen2007peerreview} generalizes this idea by 
building a practical accountable system that uses 
tamper-evident logs to detect and irrefutably identify
the faulty nodes. More recent solutions such as 
BlockchainDB~\cite{el2019blockchaindb},
BigchainDB \cite{mcconaghy2016bigchaindb}, Veritas~\cite{allen2019veritas}
and~\cite{gaetani2017blockchain} use blockchain as a 
tamper-proof log to store transactions across fully or
partially replicated databases.
CloudBFT~\cite{nogueira2014cloudbft}, on the other hand, 
tolerates malicious faults in the cloud by relying on
tamper-proof hardware to order the requests in a trusted
way.

The datastore authentication technique that uses 
Merkle Hash Trees (MHT) and Verification Objects was
first proposed by Merkle~\cite{merkle1989certified}.
The technique employed in \system that enables verifing the 
datastore per transaction is inspired by the 
work of Jain et al.~\cite{jain2013trustworthy}. 
Their solution assumes a single 
outsourced database, and more importantly, it
requires a central trusted site to store the MHT roots
of the outsourced data and the transaction history.
\system replaces the trusted entity by a 
globally replicated log that stores 
the necessary
information for authentication. Many
works have looked at query correctness, freshness, and data
provenance for static data but only few solutions such
as~\cite{li2006dynamic}
and~\cite{narasimha2006authentication} (apart from
\cite{jain2013trustworthy} discussed above) consider data updates. 
~\cite{li2006dynamic} and~\cite{narasimha2006authentication}
discuss alternate data authentication techniques but also
assume a single outsourced 
database.
\vspace{-1em}
%\vspace{-1em}
\section{Conclusion}
\label{sec:con}
Traditional data management systems typically
consider crash failures only. With the
increasing usage of the cloud, crowdsourcing, and
the rise of blockchain, the need to store data on
untrusted servers has risen. The typical approach
for achieving fault-tolerance, in general, uses
replication.  However, given the strict bounds on
consensus in malicious settings, alternative
approaches need to be explored.  In this paper, we
propose Fides, an auditable data management system designed
for infrastructures that are {\it not} trusted.
Instead of using replication for fault-tolerance,
Fides uses fault-detection to discourage malicious
behavior. An integral component of any distributed
data management system is the commit protocol. 
We propose TFCommit, a novel distributed atomic commitment protocol that executes transactions on untrusted servers.
Since every server in \system is untrusted, \system
replaces traditional transaction logs with a tamper-proof
log similar to blockchain. The tamper-proof log stores
all the necessary information required to audit the system
and detect any failures. We discuss each
component of \system i.e., the different layers of a typical
DBMS comprising of a transaction execution layer, a transaction
commitment layer, and a datastore. For each layer, both correct
execution and failure detection techniques are discussed.
To highlight the practicality of \protocol, we implement 
and evaluate \protocol.
The experiments emphasize the performance and 
scalability aspects of \protocol. 
% In conclusion,
% Fides provides a correct and solid data management system built
% on untrusted infrastructure.
% that provides a solid
% foundation for future expansions to include other 
% features of data management.

\bibliographystyle{abbrv}
\bibliography{citations.bib}

\end{document}